\PassOptionsToPackage{unicode}{hyperref}
\PassOptionsToPackage{hyphens}{url}
\PassOptionsToPackage{dvipsnames,svgnames,x11names}{xcolor}
\documentclass[
  12pt]{article}
\usepackage{amsmath,amssymb,amsthm,mathrsfs,mathtools,hyperref}
\usepackage{iftex}
\usepackage{booktabs}
\usepackage{multirow} 
\usepackage{siunitx}   % 用于对齐百分号

\usepackage{enumitem}
\ifPDFTeX
  \usepackage[T1]{fontenc}
  \usepackage[utf8]{inputenc}
  \usepackage{textcomp} % provide euro and other symbols
\else % if luatex or xetex
  \usepackage{unicode-math}
  \defaultfontfeatures{Scale=MatchLowercase}
  \defaultfontfeatures[\rmfamily]{Ligatures=TeX,Scale=1}
\fi
\usepackage{lmodern}
\ifPDFTeX\else  
\fi
\IfFileExists{upquote.sty}{\usepackage{upquote}}{}
\IfFileExists{microtype.sty}{% use microtype if available
  \usepackage[]{microtype}
  \UseMicrotypeSet[protrusion]{basicmath} % disable protrusion for tt fonts
}{}
\makeatletter
\@ifundefined{KOMAClassName}{% if non-KOMA class
  \IfFileExists{parskip.sty}{%
    \usepackage{parskip}
  }{% else
    \setlength{\parindent}{0pt}
    \setlength{\parskip}{6pt plus 2pt minus 1pt}}
}{% if KOMA class
  \KOMAoptions{parskip=half}}
\makeatother
\usepackage{xcolor}
\makeatletter
\ifx\paragraph\undefined\else
  \let\oldparagraph\paragraph
  \renewcommand{\paragraph}{
    \@ifstar
      \xxxParagraphStar
      \xxxParagraphNoStar
  }
  \newcommand{\xxxParagraphStar}[1]{\oldparagraph*{#1}\mbox{}}
  \newcommand{\xxxParagraphNoStar}[1]{\oldparagraph{#1}\mbox{}}
\fi
\ifx\subparagraph\undefined\else
  \let\oldsubparagraph\subparagraph
  \renewcommand{\subparagraph}{
    \@ifstar
      \xxxSubParagraphStar
      \xxxSubParagraphNoStar
  }
  \newcommand{\xxxSubParagraphStar}[1]{\oldsubparagraph*{#1}\mbox{}}
  \newcommand{\xxxSubParagraphNoStar}[1]{\oldsubparagraph{#1}\mbox{}}
\fi
\makeatother

\usepackage{longtable,booktabs,array}
\usepackage{calc} % for calculating minipage widths
\usepackage{etoolbox}
\makeatletter
\patchcmd\longtable{\par}{\if@noskipsec\mbox{}\fi\par}{}{}
\makeatother
\IfFileExists{footnotehyper.sty}{\usepackage{footnotehyper}}{\usepackage{footnote}}
\makesavenoteenv{longtable}
\usepackage{graphicx}
\makeatletter
\def\maxwidth{\ifdim\Gin@nat@width>\linewidth\linewidth\else\Gin@nat@width\fi}
\def\maxheight{\ifdim\Gin@nat@height>\textheight\textheight\else\Gin@nat@height\fi}
\makeatother
\setkeys{Gin}{width=\maxwidth,height=\maxheight,keepaspectratio}
\makeatletter
\def\fps@figure{htbp}
\makeatother

\makeatletter
\@ifpackageloaded{caption}{}{\usepackage{caption}}
\AtBeginDocument{%
\ifdefined\contentsname
  \renewcommand*\contentsname{Table of contents}
\else
  \newcommand\contentsname{Table of contents}
\fi
\ifdefined\listfigurename
  \renewcommand*\listfigurename{List of Figures}
\else
  \newcommand\listfigurename{List of Figures}
\fi
\ifdefined\listtablename
  \renewcommand*\listtablename{List of Tables}
\else
  \newcommand\listtablename{List of Tables}
\fi
\ifdefined\figurename
  \renewcommand*\figurename{Figure}
\else
  \newcommand\figurename{Figure}
\fi
\ifdefined\tablename
  \renewcommand*\tablename{Table}
\else
  \newcommand\tablename{Table}
\fi
}
\@ifpackageloaded{float}{}{\usepackage{float}}
\floatstyle{ruled}
\@ifundefined{c@chapter}{\newfloat{codelisting}{h}{lop}}{\newfloat{codelisting}{h}{lop}[chapter]}
\floatname{codelisting}{Listing}

\makeatother

\ifLuaTeX
  \usepackage{selnolig}  % disable illegal ligatures
\fi
\usepackage{natbib}
\usepackage{setspace}
\usepackage{subcaption}
\usepackage{bookmark}
\usepackage{algorithm}
\usepackage{algpseudocode}
\IfFileExists{xurl.sty}{\usepackage{xurl}}{} % add URL line breaks if available
\hypersetup{
  pdftitle={Title},
  pdfauthor={Author 1; Author 2},
  pdfkeywords={3 to 6 keywords, that do not appear in the title},
  colorlinks=true,
  linkcolor={blue},
  filecolor={Maroon},
  citecolor={Blue},
  urlcolor={Blue},
  pdfcreator={LaTeX via pandoc}}
\usepackage{bm}
\newtheorem{theorem}{Theorem}
\newtheorem{lemma}{Lemma}
\newtheorem{corollary}{Corollary}
\newtheorem{proposition}{Proposition}
\newtheorem{definition}{Definition}

\newtheorem{remark}{Remark}
\newtheorem{assumption}{Assumption}

\newcommand{\anon}{1}
\def\te{\widetilde{\mathbb{E}}}
\def\T{\top}
\def\supp{\mathrm{supp}}
\def\P{\mathbb{P}}
\def\te{\widetilde{\mathbb{E}}}
\newcommand{\ba}{\bm{a}}
\newcommand{\bb}{\bm{b}}

\newcommand{\bs}{\bm{s}}

\newcommand{\bw}{\bm{w}}
\newcommand{\bx}{\bm{x}}
\newcommand{\by}{\bm{y}}
\newcommand{\bz}{\bm{z}}

\newcommand{\bA}{\bm{A}}

\newcommand{\sI}{\mathscr{I}}

\newcommand{\sX}{\mathscr{X}}

\newcommand{\sZ}{\mathscr{Z}}

\newcommand{\cA}{\mathcal{A}}

\newcommand{\cE}{\mathcal{E}}

\newcommand{\cH}{\mathcal{H}}
\newcommand{\cI}{\mathcal{I}}

\newcommand{\cM}{\mathcal{M}}
\newcommand{\cN}{\mathcal{N}}

\newcommand{\cS}{{\mathcal{S}}}
\newcommand{\cT}{{\mathcal{T}}}

\newcommand{\EE}{\mathbb{E}}

\newcommand{\NN}{\mathbb{N}}
\newcommand{\PP}{\mathbb{P}}

\newcommand{\RR}{\mathbb{R}}

\newcommand{\balpha}{\bm{\alpha}}

\newcommand{\bdelta}{\bm{\delta}}
\newcommand{\bgamma}{\bm{\gamma}}

\newcommand{\btheta}{\bm{\theta}}

\newcommand{\bTheta}{\bm{\Theta}}

\newcommand{\bSigma}{\bm{\Sigma}}

\newcommand{\argmin}{\mathop{\mathrm{argmin}}}
\newcommand{\argmax}{\mathop{\mathrm{argmax}}}

\usepackage{titlesec}

\titleclass{\section}{straight} % 移除章节编号的额外间距
\titleformat{\section}
  {\normalfont\Large\bfseries}
  {\thesection}
  {1em}
  {}
  
\titlespacing*{\section}{0pt}{1ex}{0.5ex}
\titlespacing*{\subsection}{0pt}{0.8ex}{0.3ex}
\titlespacing*{\subsubsection}{0pt}{0.5ex}{0.2ex}
\titlespacing*{\paragraph}{0pt}{0.3ex}{0.1ex}

\usepackage[top=2cm, bottom=2cm, left=2.5cm, right=2.5cm]{geometry}
\usepackage{setspace}
\usepackage{lipsum}

\newlength{\mylinespacing}
\usepackage{amsmath}
\usepackage{lipsum}

\makeatletter
\g@addto@macro\normalsize{%
  \setlength\abovedisplayskip{3pt}      % 公式上方间距（原：10pt）
  \setlength\belowdisplayskip{3pt}      % 公式下方间距（原：10pt）
  \setlength\abovedisplayshortskip{0pt} % 短公式上方间距
  \setlength\belowdisplayshortskip{3pt} % 短公式下方间距
}
\makeatother

\begin{document}
\setstretch{1.0} % 重置行距因子
\setlength{\baselineskip}{\mylinespacing}

\def\spacingset#1{\renewcommand{\baselinestretch}%
{#1}\small\normalsize} \spacingset{1}

%%%%%%%%%%%%%%%%%%%%%%%%%%%%%%%%%%%%%%%%%%%%%%%%%%%%%%%%%%%%%%%%%%%%%%%%%%%%%%

\if1\anon
{
  \title{\bf Multi-Source Transfer Learning of Sparse Single-Index Models}
  \author{Ye Tian\thanks{
    Key Laboratory of Applied Statistics of MOE, Key Laboratory of Big Data Analysis of Jilin Province, School of
Mathematics and Statistics, Northeast Normal University, Changchun, Jilin 130024, P. R. China~(Email:tianye@nenu.edu.cn)}}
  \maketitle
} \fi

\if0\anon
{
  \bigskip
  \bigskip
  \bigskip
  \begin{center}
    {\LARGE\bf Multi-Source Transfer Learning of Sparse Single-Index Models}
\end{center}
  \medskip
} \fi

\bigskip
\begin{abstract}
Transfer learning leverages knowledge from related source domains to improve learning in a target domain. Recent theoretical advances cover a broad range of regression settings within (generalized) linear models. Despite their diversity, these methods share two common constraints: they assume a known link function or linear structure and require direct access to raw source data. To move beyond these constraints, we propose a source‑data‑free transfer learning framework based on the single-index model (SIM). Instead of requiring raw source data, our method transfers only summary statistics derived from a generalized Stein's lemma in a one-time communication. This design preserves privacy and avoids side effects caused by dissimilarities of unknown nonlinear link functions across domains. To capture flexible, unknown nonlinearity, we employ a multilayer perceptron guided by the pre‑estimated index from the transferred statistics, which significantly mitigates overfitting. Extensive experiments on synthetic data and a real‑world application demonstrate consistent improvements over existing (generalized) linear model‑based approaches. The proposed framework thus offers a practical, privacy‑preserving, and nonlinear‑adaptive solution for transfer learning.
\end{abstract}

\noindent%
{\it Keywords:} transfer learning, source-data-free, single index model, Stein's lemma, nonlinearity
\vfill

\newpage
\spacingset{1.8} % DON'T change the spacing!

\section{Introduction}
Transfer learning, leveraging knowledge from related sources to improve learning in a target domain, has a rich history in statistical learning~\citep{pan2010}. Recent statistical work has established rigorous theoretical foundations across a variety of regression settings.

For linear models, \citet{bastani2021predicting} consider a single‑source transfer task where the bias between the source and target parameters is assumed sparse. They propose a two‑step joint estimator and provide both $\ell_1$ and $\ell_2$ error bounds, showing that the required gold sample size can be exponentially smaller (in the dimension of the parameter) than that of naive methods. \citet{li2023transfer}  consider a multi‑source transfer problem under the assumption that a subset of source parameters are close to the target parameter in $\ell_q$ norm. They propose Oracle Trans‑Lasso (when the informative set is known) and Trans‑Lasso (when it is unknown), and demonstrate that their estimators are minimax optimal for both cases. Both works require that each informative source parameter be individually close to the target parameter in a sparse sense. In contrast, \citet{lin2024} handle a multi‑source setup under a more flexible condition: they only require that a weighted combination of the source parameters is close to the target parameter, without demanding closeness of any individual source. They develop a profiled transfer learning method and prove its minimax optimality under their assumption.

\citet{Tian02102023} extend transfer learning to high‑dimensional generalized linear models (GLMs), covering logistic, Poisson, and other families. In addition to a GLM transfer algorithm with improved estimation and prediction error bounds, they propose an algorithm‑free transferable source detection method based on cross‑validation and prove its detection consistency.

For multiple response regression, \citet{Park17102025} study transfer learning under a low‑rank structure, assuming small nuclear norms for both the target coefficient matrix and the contrasts with informative sources. Their nuclear‑norm‑penalized estimator achieves faster convergence than the single‑task reduced‑rank regression estimator. They propose a forward source detection method selecting transferable sources sequentially.

Beyond independent data, \citet{li2022transfer} study transfer learning for functional linear regression under a reproducing kernel Hilbert space framework. They propose TL‑FLR (when informative sources are known) and an aggregation‑based variant, ATL‑FLR (when informative sources are unknown), mitigating negative transfer. They further extend the results to functional generalized linear models.

\citet{zeng2026transfer} propose a transfer learning framework for spatial autoregressive models to handle spatially dependent data. They develop a tranSAR algorithm and introduce a transferable source detection method based on spatial residual bootstrap preserving spatial dependence. Theoretical convergence rates and detection consistency are established.

Despite the significant theoretical progress, two major limitations persist in the existing statistical literature on regression‑oriented transfer learning. 

First, linear models have been extensively analyzed~\citep{bastani2021predicting,li2022transfer,lin2024}, and beyond linearity, generalized linear models (GLMs) represent the primary extension studied in transfer learning~\citep{Tian02102023}. However, GLMs impose known link functions. While this captures certain nonlinearities, it does not handle the more general setting where the nonlinear transformation is completely unknown and must be learned from data. Consequently, existing theoretical work lacks a treatment of flexible, unknown nonlinearity, especially when different domains pose various nonlinear link functions, a situation that arises naturally in many real-world applications. Transferring knowledge under such a setup introduces at least two problems that differ from the linear case. First, how to handle the unknown nonlinearity; second, how to define transferability in this context. 

Moreover, nearly all existing statistical methods assume direct and simultaneous access to both source and target data. This assumption is increasingly impractical, as source domain data may be inaccessible due to various real‑world constraints, for example, data protection regulations, proprietary restrictions, logistical costs, or the simple fact that the source model has already been trained and the raw data can no longer be shared. Consequently, existing methods offer little guidance for transfer learning when only summary information or pre‑trained models are accessible.

In this work, we address both limitations via the single‑index model (SIM). We exploit similarity among indices across domains, which in fact reflects similarity in the one‑dimensional embedding spaces defined by those indices. To avoid accessing raw source data, we transfer only the estimated source indices derived from a generalized Stein's lemma, a one‑time communication of summary statistics that requires no direct exposure to the raw source data. This is particularly advantageous when the raw source data cannot be shared due to practical constraints. To capture flexible, unknown nonlinearity, we then employ a multilayer perceptron (MLP) guided by the pre‑estimated indices obtained from these transferred statistics, which significantly mitigates overfitting. This dual design ensures both nonlinear flexibility and practical viability under common data‑sharing limitations.

\textbf{Notations.} Vectors are denoted by bold lowercase letters (e.g., $\ba, \bb$). Matrices are denoted by bold uppercase letters. For a vector $\ba \in \RR^d$ and $j \in [d]_+$, let $[\ba]_j$ denote the $j$-th coordinate; $\ba_{m:n}$ denote the contiguous slice from index $m$ to $n$ ($1 \leq m \leq n \leq d$); $[\ba]_{\cI}$ denote the slice over an index set $\cI \subseteq [d]_+$; $\cS^c$ denote the complement of a set $\cS$ within $[d]_+$, i.e., $\cS^c \triangleq [d]_+ \setminus \cS$; $\supp(\ba) \triangleq \{j \in [d]_+ : [\ba]_j \ne 0\}$ denote the support of $\ba$. A vector is \textit{$s$-sparse} if $|\supp(\ba)| = s$. For $k \in \NN_+$, $[k]_+ \triangleq \{1, \dots, k\}$. For $k \in \NN$, $[k] \triangleq \{0, 1, \dots, k\}$.  Given a set of samples $\cS = \{\bz_i\}_{i=1}^n \subset \sZ$ and a function $f: \sZ \to \RR$, the empirical average is
$\widetilde{\mathbb{E}}_{\cS}[f(\bz)] \triangleq (1/n) \sum_{i=1}^n f(\bz_i)$. When $f$ is the identity, we write $\widetilde{\mathbb{E}}_{\cS}[\bz]$. The subscript $\cS$ may be omitted if clear. For two sequences $\{a_n\}$ and $\{b_n\}$, $a_n = o(b_n)$ iff $\displaystyle \lim_{n \to \infty} (a_n/b_n) = 0$; $a_n \asymp b_n$ iff there exist constants $C_1, C_2 > 0$ such that $C_1 |b_n| \leq |a_n| \leq C_2 |b_n|$ for all sufficiently large $n$. For a scalar $\tau > 0$ and a vector $\ba \in \RR^d$, the element-wise hard thresholding operator $\mathrm{HT}(\ba, \tau)$ is defined as $
\mathrm{HT}(\ba, \tau) = \left\{ \mathrm{H}([\ba]_1, \tau), \dots, \mathrm{H}([\ba]_d, \tau) \right\}^{\T}$,    where the scalar hard thresholding function is
$\mathrm{H}(c, \tau) = c$, if $|c| > \tau$; and $0$, otherwise.  
\section{Preliminaries}

\subsection{the Model}
In this work, we consider the following multi-source transfer learning framework of the SIM.
Specifically, we have $K$ source domains demoted by $i \in [K]_+$ and one target domain denoted by $i = 0$. Suppose i.i.d. observations from the $i$-th  domain, $\{(\bx_j^{(i)}, \by_j^{(i)})\}_{j=1}^{n_{i}}$ satisfy the SIM of the corresponding domain: 
\begin{align}\label{eq:sim}
    y^{(i)} = f_{i}(\balpha^{\top}_{i}\bx^{(i)}) + \epsilon_{i}.
\end{align}

In Model~\eqref{eq:sim}, $\balpha_0$ is the target index of interest and $\balpha_i$ ($i \in [K]_+$) are source indices  encoding transferable information. They share the same dimension $d$. The model is flexible: covariates $\bx^{(i)}$'s may be distributed differently; noises $\epsilon_i$'s independent from covariates could follow distinct distributions; the unknown nonlinear link functions $f_i$'s can also vary across domains. 

Without further constraints, $\balpha_i$ is not identifiable given an arbitrary $f_i$. Following the standard SIM convention, we assume $\|\balpha_i\|_2 = 1$ for all $i \in [K]$, so that $\balpha_i$ is identifiable up to the sign. We further fix the sign by requiring $\mathbb{E}\{ \nabla f_i(\balpha_i^{\top}\bx^{(i)}) \} $ to be positive. Moreover, we assume $\balpha_i$'s are $s_i$-sparse, respectively. For $\balpha_i$, we define the corresponding scale and scaled parameter as
$\mu_i = \mathbb{E}\{ \nabla f_{i} (\balpha_{i}^{\top}\bx^{(i)})\}$ and $ \widetilde{\balpha}_{i} = \mu_i \balpha_i$. In general, to estimate the normalized parameters $\balpha_i$'s, we first need to estimate their scaled correspondences $\Tilde{\balpha}_i$'s.  
For each source domain $i \in [K]_+$, the contrast vector is defined by $\bdelta_i = \balpha_i - \balpha_0$. To enable knowledge transfer, $\bdelta_i$'s are required to satisfy certain sparse conditions formally stated in Section~\ref{sec:theory}.

\subsection{the Stein's Score and Generalized Stein's Lemma}

Our estimators of $\widetilde{\balpha}_i$ will be constructed by Stein's score and a generalized Stein’s lemma. We first introduce these technical tools below. 

\begin{definition}[Stein's Score]
Let $\bx \in \mathbb{R}^{d}$ be a random vector with density $\P(\bx)$. If $\P(\bx)$ is differentiable, the first-order score function, denoted by $\bs(\bx): \mathbb{R}^{d} \rightarrow \mathbb{R}^{d}$, is defined as
\begin{align}\label{eq:def-fs}
    \bs(\bx) \coloneqq - \nabla_{\bx}[\ln\{\PP(\bx)\}] = - \nabla_{\bx}  \PP(\bx)  / \PP(\bx).
\end{align}
\end{definition}

\begin{lemma}[Generalized Stein's Lemma]\label{lem:fs}
Suppose Model~\eqref{eq:sim} holds. Assume that  the expectations $\mathbb{E}\{y^{(i)}\bs(\bx^{(i)})\}$ and $\mathbb{E}\{\nabla f_{i}(\balpha_i^{\top}\bx^{(i)})\}$ both exist and well-defined for $i \in [K]$. Further assume that $\lim_{\| \bx^{(i)}\| \rightarrow \infty }f_{i}(\balpha_i^{\top}\bx^{(i)}) \P(\bx^{(i)}) \rightarrow 0$.  Then, $\mathbb{E}\{ y^{(i)} \bs(\bx^{(i)})\} = \widetilde{\balpha}_i$ holds.  
\end{lemma}

Given the true score function $\bs(\bx^{(i)})$, the moment estimator for the scaled parameter $\widetilde{\balpha}_i$ is defined as $\cM_{i} = \te\{ y^{(i)} \bs(\bx^{(i)})\}$, per Lemma~\ref{lem:fs}. 

\subsection{Score Estimators}

In the literature on Stein’s method, the distributions of inputs are conventionally assumed to be known \citep{2017zhuoran,balasubramanian2018,na2019high}; consequently, the score functions are available and often analytically tractable.

In recent years, with the rise of deep learning and diffusion models in generative AI, deep neural networks (DNN) have been widely used to estimate Stein’s scores. Despite their powerful approximation capabilities, existing error analyses focus primarily on the mean squared error~\citep{chen2023score,shen2024}, the approximation error ~\citep{yakovlev2025a}, or on the divergence of distributions~\citep{oko2023diffusion,yakovlev2025b}. To our knowledge, there is no relevant analysis of pointwise error bounds for neural network score estimators that would facilitate end-to-end error analysis when these estimators are used as an intermediate step.

In contrast, kernel methods previously prevalent in score function estimation \citep{strathmann2015gradient,li2018gradient,shi2018spectral,zhou2020} can provide such bounds under certain regularity conditions. Since point evaluation functionals in the RKHS are bounded linear functionals, pointwise errors can be uniformly bounded by the RKHS norm. \citet{zhou2020} provide an error bound in terms of the RKHS norm, and we derive the corresponding pointwise error rate as a corollary of their Theorem $\mathrm{B}.1$. The details are deferred to Section $\mathrm{I}.2$ of the supplement. Roughly speaking, for their non-parametric method on the input domain $\sX$, we have
\[
\sup_{\bx \in \sX} \|\widehat{\bs}(\bx) - \bs(\bx)\|_2 = O\!\left\{M^{-\frac{r}{2(r+1)}}\right\},
\]
where $M$ is the sample size used to estimate the score function and $r$ is the source condition index that measures how well the true score function's smoothness aligns with the RKHS of the kernel.

Compared with DNN methods, kernel methods are theoretically clean but computationally heavy for large sample sizes. We will demonstrate the performance of the former in Section~\ref{sec:ss} with large samples, and that of the latter in Section~\ref{sec:rda} with relatively small ones.

In this work, we make the following assumption about the score estimator, which holds for several non-parametric score estimators~\citep{zhou2020} as well as DNN estimators such as those in~\citet{chen2023score} and \citet{shen2024}.

\begin{assumption}\label{ass:sf}
For any $i \in [K]$, suppose $\widehat{\bs}_i(\cdot)$ is estimated on $n_{s,i}$ i.i.d. samples on the $i$-th domain, there exists a decreasing function  $\phi(\cdot)$ going to 0 as $n_{s,i}$ goes to infinity, such that $(\EE[\{\widehat{\bs}_i(\bx^{(i)}) - \bs(\bx^{(i)})\}^2])^{1/2} \leq \phi(n_{s,i})$.   
\end{assumption}

Moreover, given a score estimator $\widehat{\bs}_i(\cdot)$ for a specific domain $i$, we use $\widehat{\cM}_i$ to denote $\te\{ y^{(i)} \widehat{\bs}_i(\bx^{(i)})\}$ in the following.

\section{Methodology}

In this section, we introduce the proposed methods of the transfer learning framework. We introduce the normalized hard‑thresholding estimator of $\balpha_i$ and the baseline estimator of $\balpha_0$ as the intermediate estimators in Sections~\ref{sec:sp} and~\ref{sec:asi}, respectively. The simple average estimator (SAE) and the optimal convex combination estimator (OCCE) of $\balpha_0$ are introduced in Section~\ref{sec:etp}. In Section~\ref{sec:ssp}, we present a procedure for informative source selection. Finally, an MLP with a single hidden layer estimating the target link function is described in Section~\ref{sec:elf}. Theoretical values of all tuning parameters are given in Section~\ref{sec:theory}, and in practice they are chosen by cross‑validation; details are deferred to Section $\mathrm{II}.2$ of the supplement.     

\subsection{Estimates of Source Parameters}\label{sec:sp}

After obtaining the moment estimator $\widehat{\cM}_i$, the scaled parameter $\widetilde{\balpha}_{i}$ is estimated on the $i$-th source domain by solving the following optimization problem:
\begin{align}\label{eq:obj}
\balpha_i^{\dagger} =
\argmin_{\balpha \in \RR^d} \| \balpha - \widehat{\cM}_i \|^{2}_{2} + \lambda^2_i\| \balpha \|_0. 
\end{align}
The solution to Problem~\eqref{eq:obj} is the hard-thresholding estimator,
\begin{equation*}
\balpha_i^{\dagger} = \mathrm{HT}[\widehat{\cM}_i, \lambda_i],
\end{equation*}
i.e., each component of $\widehat{\cM}_i$ is kept if its absolute value exceeds $\lambda_i$ and set to zero otherwise. The final estimator of $\balpha_i$ is then the normalized hard-thresholding estimator $\widehat{\balpha}_i = \balpha_i^{\dagger} / \|\balpha_i^{\dagger}\|_2$, which serves as the only information shared from the $i$-th source domain in the entire transfer learning process.

\subsection{Baseline Estimator of the Target Parameters }\label{sec:asi}

On the target domain, due to its limited observations, a more robust estimator is favored. Therefore, we propose to estimate $\widetilde{\balpha}_0$ via Lasso:
\begin{align}\label{eq:obt}
\balpha_0^{\dagger} = \argmin_{\balpha \in \mathbb{R}^d} \frac{1}{2}\| \widehat{\cM}_{0} - \balpha_0 \|^2_2 + \lambda_0 \| \balpha_0 \|_1.
\end{align}
The solution to Problem~\eqref{eq:obt} is simply the soft-thresholding estimator:
\begin{equation*}
\balpha_0^{\dagger} = \mathrm{ST}[\widehat{\cM}_0, \lambda_0],
\end{equation*}

and the normalized soft-thresholding estimator $\widehat{\balpha}_0^{\text{base}} = \balpha_0^{\dagger} / \|\balpha_0^{\dagger}\|_2$ would serve as the baseline estimator of $\balpha_0$.
\subsection{Estimates of the Target Parameters}\label{sec:etp}
Let $\bar \bdelta = (1/K) \sum_{i=1}^{K} \bdelta_{i}$, if $K$ is relatively small or the supports of $\bdelta_{i}$'s overlap sufficiently, then, $\bar \bdelta$  remains sufficiently sparse. In this case, we propose to use the following simple average estimator.

\subsubsection{the Simple Average Estimator}

Let $\balpha_{s} = (1/K)\sum_{i=1}^{K} \balpha_i$, then, it can be estimated by $\widehat{\balpha}_s = (1/K)\sum_{i=1}^{K} \widehat{\balpha}_{i}$. And $\bar \bdelta$ can be estimated by Lasso: 
\begin{align}\label{eq:obd}
    \widehat\bdelta = \argmin_{\bdelta \in \RR^{d}} \frac{1}{2}\| \widehat{\balpha}_s - \widehat{\balpha}_0^{\text{base}} - \bdelta \|_2^2 +  \gamma_{\bar \bdelta}\| \bdelta \|_1.
\end{align}
Similar to Problem~\eqref{eq:obt}, the solution to Problem~\eqref{eq:obd} is the soft thresholding estimator:
\begin{align*}
    \widehat\bdelta = \mathrm{ST}[\widehat{\balpha}_s - \widehat{\balpha}_0^{\text{base}},  \gamma_{\bar \bdelta}]
\end{align*}

Let $\balpha^{\dagger}_{0,s} = \widehat{\balpha}_s - \widehat{\bdelta}$, then, $\balpha_0$ can be estimated by normalizing $\balpha^{\dagger}_{0,s}$:
\begin{align*}
\widehat{\balpha}^{\text{sae}}_{0} = \frac{\balpha^{\dagger}_{0,s}}{\| \balpha^{\dagger}_{0,s} \|_2}.  
\end{align*}

We refer to
$\widehat{\balpha}^{\text{sae}}_{0}$ as simple average estimator (SAE), since it originates from a simple average of source estimators. The procedure is outlined in Algorithm~\ref{alg:sae}.

\begin{algorithm}
\caption{Simple Average Estimator of $\balpha_0$}
\label{alg:sae}
\begin{algorithmic}[1]
\Require  Normalized hard-thresholding estimators, $\{\widehat{\balpha}_i\}_{i=1}^K$, baseline estimator $\widehat{\balpha}_0^{\text{base}}$ and the hyper-parameter $\gamma_{\bar \bdelta}$.
\Ensure Simple average estimator $\widehat{\balpha}^{\text{sae}}_{0}$.
\State $\widehat{\balpha}_s \gets (1/K) \sum_{i=1}^{K} \widehat{\balpha}_{i}$ 
\State $ \widehat\bdelta \gets \mathrm{ST}[\widehat{\balpha}_s - \widehat{\balpha}_0^{\text{base}}, \gamma_{\bar \bdelta}]$ 
\State $\balpha^{\dagger}_{0,s} \gets \widehat{\balpha}_s - \widehat{\bdelta}$
\State $\widehat{\balpha}^{\text{sae}}_{0} \gets \balpha^{\dagger}_{0,s} / \| \balpha^{\dagger}_{0,s}
 \|_2$
\end{algorithmic}
\end{algorithm} 

If $K$ is relatively large or the sources are not similar enough, $\bar \bdelta$ may lack sufficient sparsity. In such scenarios, we propose the following optimal convex combination estimator.

\subsubsection{the Optimal Convex Combination Estimator}

Similar to $\bar \bdelta$, for $i \in [K]_+$, $\bdelta_i$ can be estimated by Lasso: 
\begin{align*}
    \widehat\bdelta_i = \argmin_{\bdelta \in \RR^{d}} \frac{1}{2}\| \widehat{\balpha}_i - \widehat{\balpha}_0^{\text{base}} - \bdelta \|_2^2 + \lambda_{\bdelta_i}\| \bdelta \|_1,
\end{align*}
and $\widehat\bdelta_i = \mathrm{ST}[\widehat{\balpha}_i - \widehat{\balpha}_0^{\text{base}}, \lambda_{\bdelta_i}]$. Consequently, $\balpha_0$ can be estimated by:
\begin{align*}
\widehat{\balpha}_{0,i} = \frac{\balpha^{\dagger}_{0,i}}{\| \balpha^{\dagger}_{0,i} \|_2}, 
\end{align*}
where $\balpha^{\dagger}_{0,i} = \widehat{\balpha}_i - \widehat{\bdelta_i}$.
 
The collection $\cH = \{\widehat{\balpha}_{0,1}, \ldots, \widehat{\balpha}_{0,K} \}$ thus forms a dictionary for estimating $\balpha_0$.

For any weight vector $\bgamma = ([\bgamma]_1, \ldots, [\bgamma]_K)^{\top}$ in the $(K-1)$-dimensional simplex $\Delta^{K-1} \coloneqq \{ \bgamma \in \mathbb{R}^K : \gamma_i \ge 0, \sum_{i=1}^K \gamma_i = 1 \}$, let $\balpha^{\dagger}_{0, \bgamma} = \sum_{i=1}^{K} [\widetilde{\bgamma}]_i \widehat{\balpha}_{0,i}$, 
define the convex combination estimator as  $\widehat{\balpha}_{0, \bgamma} = \sum_{i=1}^{K} \bgamma_i \balpha^{\dagger}_{0, \bgamma}/\left\| \balpha^{\dagger}_{0, \bgamma} \right\|_2$. Then, the optimal weight $\widetilde{\bgamma}$ can be obtained by solving:
\begin{align*}
\widetilde{\bgamma} = \argmin_{\bgamma \in \Delta^{K-1}} \left\| \widehat{\balpha}_{0, \bgamma} -  \widehat{\balpha}^{\text{base}}_{0}\right\|_2^2.
\end{align*}
We then define the optimal convex combination estimator (OCCE) as $\widehat{\balpha}^{\text{occ}}_{0} = \widehat{\balpha}_{0, \widetilde{\bgamma}}$. The complete procedure is summarized in Algorithm~\ref{alg:gamma_combination}.
\begin{algorithm}[ht]
\caption{Optimal Convex Combination Estimator of $\balpha_0$}
\label{alg:gamma_combination}
\begin{algorithmic}[1]
\Require Normalized hard-thresholding estimators $\{\widehat{\balpha}_i\}_{i=1}^K$, hyper-parameters $\{\gamma_{\bdelta_i}\}_{i=1}^K$, and the base estimator $\widehat{\balpha}_0^{\text{base}}$.
\Ensure Optimal convex combination estimator $\widehat{\balpha}^{\text{occ}}_{0}$.
\Statex
\For{$i = 1$ to $K$}
    \State $\widehat\bdelta_i \gets \mathrm{ST}[\widehat{\balpha}_i - \widehat{\balpha}_0^{\text{base}}, \gamma_{\bdelta_i}]$ 
    \State $\balpha^{\dagger}_{0,i} \gets \widehat{\balpha}_i - \widehat{\bdelta_i}$  
    \State $\widehat{\balpha}_{0,i} \gets \balpha^{\dagger}_{0,i} / \|\balpha^{\dagger}_{0,i} \|_2$ 
\EndFor

\State Solve for optimal weights:
\[
\widetilde{\bgamma} = \argmin_{\bgamma \in \Delta^{K-1}} F(\bgamma)
\]
\State $\balpha^{\dagger}_{0, \widetilde{\bgamma}} \gets \sum_{k=1}^{K} \widetilde{\bgamma}_k \widehat{\balpha}_{0,k}$ 
\State $\widehat{\balpha}^{\text{occ}}_{0} \gets\balpha^{\dagger}_{0, \widetilde{\bgamma}}/ \|\balpha^{\dagger}_{0, \widetilde{\bgamma}}\|_2$ 
\end{algorithmic}
\end{algorithm}

Since both $\widehat{\balpha}^{\text{base}}_{0}$ and $\widehat{\balpha}_{0, \bgamma}$ are unit vectors, minimizing $\left\| \widehat{\balpha}_{0, \bgamma} - \widehat{\balpha}^{\text{base}}_{0}\right\|_2^2$ is equivalent to maximizing their correlation:
\begin{align*}
    F(\bgamma) =  (\widehat{\balpha}^{\text{base}}_{0})^{\T} \widehat{\balpha}_{0, \bgamma} = \frac{(\widehat{\balpha}^{\text{base}}_{0})^{\T}\sum_{i=1}^{K} \bgamma_i \widehat{\balpha}_{0,i}}{\left\| \sum_{k=1}^{K} \bgamma_i \widehat{\balpha}_{0,i} \right\|_2}.
\end{align*}
Defining the vector $\bb \in \RR^{K}$ with $\bb_i = (\widehat{\balpha}^{\text{base}}_{0})^{\T}\widehat{\balpha}_{0,i}$ and the Gram Matrix $\bA \in \RR^{K \times K}$ with $\bA_{i,j} = \widehat{\balpha}^{\T}_{0,i}\widehat{\balpha}_{0,j}$, we can rewrite $F(\bgamma)$ as
$F(\bgamma) = \bb^{\T}\bgamma/\sqrt{\bgamma^{\T}\bA \bgamma}$.
Therefore, the optimization problem reduces to
$\widetilde{\bgamma} = \argmax_{\bgamma \in \Delta^{K-1}} F(\bgamma)$,
which can be efficiently solved by the Frank-Wolfe algorithm, as detailed in Algorithm 1 in Section $\mathrm{II}.1$ of Supplement.

\subsection{the Source Selection Procedure}\label{sec:ssp}

Algorithms~\ref{alg:sae} and~\ref{alg:gamma_combination} require prior knowledge of which sources are informative (this concept is defined in Section~\ref{sec:fa}). In practice, however, this information is often unavailable. Transferring uninformative sources may be unhelpful or even degrade performance compared to using the target data alone, a phenomenon known as negative transfer~\citep{pan2010,torrey2010transfer,weiss2016survey,li2022transfer,Tian02102023}.

To avoid this problem, we propose a source selection procedure (Algorithm~\ref{alg:source_selection}) based on estimated contrasts. Under certain assumptions, this procedure correctly identifies informative versus non‑informative sources with high probability; its consistency of selection is formally established in Proposition~\ref{prop:s}.
\begin{algorithm}[htbp]
\caption{Selection of Informative Sources by Infinity Norm Ranking}
\label{alg:source_selection}
\begin{algorithmic}[1]
\Require Baseline estimator $\widehat{\boldsymbol{\alpha}}_0^{\text{base}}$, normalized hard-thresholding estimators $\{\widehat{\boldsymbol{\alpha}}_i\}_{i=1}^K$, number of desired informative sources $k$.
\Ensure The set of informative sources $\mathcal{I}$ (size $k$)
\For{$i = 1$ to $K$}
    \State $\bdelta^{\dagger}_i \gets \widehat{\boldsymbol{\alpha}}_i - \widehat{\boldsymbol{\alpha}}_0^{\text{base}}$
    \State $r_i \gets \|\bdelta^{\dagger}_i \|_{\infty}$
\EndFor
\State Sort indices by $r_i$ in ascending order to obtain $\text{rank}(1), \dots, \text{rank}(K)$
\State $\mathcal{I} \gets \{\, \text{rank}(1), \dots, \text{rank}(k) \,\}$
\end{algorithmic}
\end{algorithm}

In our setting, two factors necessitate prior knowledge of the number of informative sources or the use of conservative source‑selection strategies. First, methods such as that of \citet{Tian02102023} rely on access to source data, which is unavailable in our source‑data‑free framework; their detection procedure therefore cannot be applied. Second, due to the presence of unknown nonlinear link functions in the target domain, simple surrogate losses like MSE, which are friendly to small sample sizes under linear models, cannot reliably guide automatic threshold selection via cross‑validation. We illustrate this second issue using the method of \citet{Tian02102023} as an example in Section~\ref{sec:exp-ssa}.

\subsection{the Estimate of $f_0$}\label{sec:elf}
We estimate $f_0$ with a single-hidden-layer neural network: $\widehat{f}_{\btheta}(x) = \bw_o^{\top}\phi(\bw_h x + \bb_{h}) + b_o$, where $\btheta = (b_o, \bb_h^{\top}, \bw_o^{\top}, \bw_h^{\top})^{\top}$ comprises all learnable parameters and $\phi$ is an almost-everywhere differentiable activation function. Given an estimator $\widehat{\balpha}_0$ of $\balpha_0$, we fit $\btheta$ by solving
\begin{align*}
\widehat{\btheta}_{\text{e}}(\widehat{\balpha}_0) = \argmin_{\btheta \in \bTheta} \te_{\cT_{f_0}} [\{ y - \widehat{f}_{\btheta}(\widehat{\balpha}_{0}^{\T}\bx) \}^2],    
\end{align*}
where $\bTheta \subset \mathbb{R}^d$ is the parameter space and $\cT_{f_0}$ is the training set. While $\cT_{f_0}$ could be a subset of $\cT_0$ for theoretical convenience, we use $\cT_{f_0} = \cT_0$ to maximize data efficiency, which does not affect the analysis in Section~\ref{sec:ed}.

For notational convenience, we abbreviate $\widehat{\btheta}_{\text{e}}(\widehat{\balpha}_0)$ as $\widehat{\btheta}_{\text{e}}$ when no confusion arises. Correspondingly, we denote the estimated link function based on $\widehat{\balpha}_0$ as $\widehat{f}_{\widehat{\balpha}_0} \coloneqq \widehat{f}_{\widehat{\btheta}_{\text{e}}}$. Finally, the composed estimator of the target function $f_0\{\balpha_0^{\top} (\cdot)\}$ is denoted by $g_{\widehat{\balpha}_0} \coloneqq \widehat{f}_{\widehat{\balpha}_0} \{\widehat{\balpha}^{\top}_0 (\cdot)\}$.

\section{Theoretical Analysis}\label{sec:theory}
This section provides theoretical analyses of the proposed methods. We begin by outlining the fundamental definition and assumptions in Section~\ref{sec:fa}. Section~\ref{sec:hte} establishes the convergence properties of the normalized hard-thresholding estimators, and Section~\ref{sec:be} does the same for the baseline estimator. Under the assumption that all sources are informative, we then derive the convergence rate of the simple average estimator (Section~\ref{sec:psae}) and analyze the optimal convex combination estimator (Section~\ref{sec:occe}). Furthermore, Section~\ref{sec:issp} proves the selection consistency of our informative-source selection procedure under mild conditions. Finally, Section~\ref{sec:ed} provides an error decomposition of the composed estimator when both the target index and the link function are estimated. All proofs are deferred to the supplementary materials.
\subsection{Fundamental definiton and Assumptions}\label{sec:fa}

We begin by stating the fundamental definition and assumptions governing our setup and underpinning the possibility of positive knowledge transfer.

\begin{definition}[Informative Source]\label{def:inf} 
Let $h_i = \| \bdelta_i \|_1$, we say the $i$-th source is informative if $s_0 \asymp s_{i}$ and $h_i = o\left\{ \sqrt{s_0\ln(2n_{0}^2)/n_{0}}+ \sqrt{s_0}\phi(n_0)\right\}$.
\end{definition}

Our requirements for informative sources are two‑fold. First, the source indices  should be as sparse as the target index, so that the two are similar in sparsity pattern. Second, their contrasts need to be small enough to allow the proposed methods to exploit the similarity (this is where the $\phi(n_0)$ term arises). Here we impose constraints on $\|\bdelta_i\|_1$ rather than $\|\bdelta_i\|_0$, since this is more realistic: in practice, small differences may exist in certain coordinates. In the following, unless stated otherwise, we assume that all sources in $[K]_+$ are informative.       

\begin{assumption}\label{ass:order} 
For all $j, k \in [K]_+$, $s_j \asymp s_k$, $h_j \asymp h_k$ and $n_j \asymp n_k$. Furthermore, $n_{0} = o(n_{j})$,  and $s_0/\sqrt{n_{0}/\ln(2n_{0}^2)} = o(1)$.   
\end{assumption}
 
\begin{assumption}~\label{ass:d}
As $n_0 \rightarrow \infty$, $d/(n_0^2) \rightarrow 0$. 
\end{assumption}
Assumption~\ref{ass:order} specifies a common transfer setting: sample-rich source domains contrasted with a data-scarce target domain, as often encountered with underrepresented groups or new system users. Moreover, it requires the level of information from all sources to be the similar; otherwise, the less informative sources can be eliminated. It further requires the indices to be sufficiently sparse and to scale appropriately with the sample sizes. Assumption~\ref{ass:d} controls the regime of dimensionality we consider for the target domain. Since $\mu_i$'s are no longer treated as fixed, to ensure that Model~\eqref{eq:sim} remains well-defined, we introduce the following assumption.
\begin{assumption}~\label{ass:mu}
For any $i \in [K]$, $\exists \; c_{i,0} > 0$. such that $|\mu_i| > c_{i,0}$ as $n_0 \rightarrow \infty$.      
\end{assumption}

\subsection{Properties of the Normalized Hard Thresholding Estimators}\label{sec:hte}

The derivation of the hard thresholding estimator's properties requires a set of assumptions on the underlying score and link functions, which we specify below. 

\begin{assumption}\label{ass:score}
For any $i \in [K]$, and any $j \in [d]_+$, $[\widehat{\bs}_i(\bx^{(i)})]_{j}$ is sub-Gaussian, i.e., $\exists \, c_{i,1} > 0$ a constant, s.t. $\sup_{\ell \in \NN_{+}}\ell^{-1/2}[\EE\{|[\widehat{\bs}_i(\bx^{(i)})]_{j}|^{\ell}\}]^{1/\ell} \leq c_{i,1}$. 
\end{assumption}

Assumption~\ref{ass:score} requires the score estimators to be element-wise sub-Gaussian, a condition commonly imposed on the true score function in the related literature~\citep{db2018}. For kernel score estimators with point-wise convergence guarantees~\citep{zhou2020}, sub-Gaussianity is preserved if the true score function is sub-Gaussian. However, whether DNN score estimators preserve sub-Gaussianity is case‑dependent and may require further constraints.

Nevertheless, in principle, the sub-Gaussian assumption can be relaxed to finite moment conditions, which is sufficient for applying Stein’s identity. \citet{2017zhuoran}, for instance, required a finite fourth moment via truncation. While this suggests a viable approach to handle heavy-tailed scores, such extensions are not central to the current analysis.

\begin{assumption}\label{ass:lf} 
For any $i \in [K]$, $f_{i}(\balpha^{\T}_{i}\bx^{(i)})$ is sub-Gaussian, i.e.,  $\exists \, c_{i,2} >0$ a constant, s.t.
 $\sup_{\ell \in \NN_{+}} \ell^{-1/2} [\EE\{ 
   f_{i}(\balpha^{\T}_{i}\bx^{(i)})^{\ell}\}]^{1/\ell} \leq c_{i,2}$ . 
\end{assumption}

Assumption~\ref{ass:lf} further constrains the behavior of the link functions, a condition satisfied by many common function classes including Lipschitz functions.

Under these assumptions, we establish Lemma~\ref{lem:sbu-g}, which characterizes the convergence rate of
the moment estimator.

\begin{lemma}\label{lem:sbu-g}
Under the condition of Lemma~\ref{lem:fs},   Assumptions~\ref{ass:sf},~\ref{ass:score}, and~\ref{ass:lf}. For any $i \in [K]$, if $\epsilon_{i}$ in model~\eqref{eq:sim} follows sub-gaussian distributions with mean 0, then,  $\exists \; c_{i,3}$'s depending on the sources of data, such that for any $j \in [d]_+$, with probability at most $\delta$,  
\begin{align}
 |[\widehat{\cM}_{i}]_j - [\widetilde{\balpha}_i]_j|\geq c_{i,3} \sqrt{\ln(2/\delta)/n_i} + \sqrt{2} c_{i,2} \phi(n_i).
\end{align}
\end{lemma} 
In the following, let $ \cE_{i,j}$ denote the event $\{|[\widehat{\cM}_i]_j - [\widetilde{\balpha}_i]_j| > \lambda_{i}\}$, where $\lambda_{i} = \cdot \left \{ c_{i,3} \sqrt{\ln(2n_{i}^2)/n_{i}}+ \sqrt{2} c_{i,2} \phi(n_i)\right \}$ for $i \in [K]_+$ and $\lambda_{0} = 2 \cdot \left \{ c_{0,3} \sqrt{\ln(2n_{0}^2)/n_{0}}+ \sqrt{2} c_{0,2} \phi(n_0)\right \}$. Apparently, $\P(\cE_{i,j}) \leq \delta$. 

Building on Lemma~\ref{lem:sbu-g}, we can derive the following properties of the normalized hard thresholding estimator $\widehat{\balpha}_{i}$.

\begin{proposition}
\label{prop:bs-bound}
Under the conditions of Lemma~\ref{lem:sbu-g}, for any $i \in [K]_+$,  for large enough $n_i$, on the event $\left(\cup_{j \in [d]_+} \cE_{i,j}\right)^c$, the following holds: 
\begin{subequations}
 \begin{align}\label{eq:supp-bd-1}
\|\widehat{\balpha}_i - \balpha_i\|_1 \leq \frac{4s_{i} \lambda_i}{c_{i,0} - 2 \sqrt{s_{i}} \lambda_{i}} ,
\end{align} 
\begin{align}\label{eq:supp-bd-2}
    \|\widehat{\balpha}_i - \balpha_i \|_2 \leq \frac{4\sqrt{s_i} \lambda_i}{c_{i,0} - 2 \sqrt{s_{i}} \lambda_{i}},  
\end{align}
and
\begin{align}\label{eq:supp-bd-2}
   \|\widehat{\balpha}_i - \balpha_i \|_{\infty} \leq \frac{2 \lambda_{i}}{c_{i,0} - 2 \sqrt{s_{i}} \lambda_{i}}  + \frac{2 \sqrt{s_{i}} \lambda_{i}}{c_{i,0} - 2 \sqrt{s_{i}} \lambda_{i}}.  
\end{align}
\end{subequations}      
\end{proposition}

In the following, for $i \in [K]_+$, let $\gamma_i = 2 \lambda_{i}/(c_{i,0} - 2 \sqrt{s_{i}} \lambda_{i})  + 2 \sqrt{s_{i}} \lambda_{i}/(c_{i,0} - 2 \sqrt{s_{i}} \lambda_{i})$.

\subsection{Properties of the Baseline Estimator}\label{sec:be}
By Lemma~\ref{lem:sbu-g}, we can derive the following properties of the baseline estimator.
\begin{proposition}\label{prop:baseline}
Under the conditions of Lemma~\ref{lem:sbu-g}, for large enough $n_0$ so that $3\sqrt{s_{0}} \lambda_0 < 2c_{0,0}$,  on the event $\left(\cup_{j \in [d]_+} \cE_{0,j}\right)^c$, the following holds: 

\begin{align}\label{eq:saeb0}
 \left \|\balpha^{\dagger}_{0} - \widetilde{\balpha}_{0} \right \|_1 \leq 6 s_0 \lambda_0
 \text{ and }
 \left \| \balpha^{\dagger}_{0} - \widetilde{\balpha}_{0} \right \|_2 \leq \frac{3}{2} \sqrt{s_0}\lambda_0.     
 \end{align}
Moreover, 
\begin{subequations}
\begin{align}\label{eq:saebb0}
 \left \|\widehat{\balpha}^{\text{base}}_{0} - \balpha_{0} \right \|_1 \leq \frac{15s_0\lambda_0}{2c_{0,0} - 3\sqrt{s_{0}} \lambda_0},    
 \end{align}
 \begin{align}\label{eq:saebb1}
 \left \| \widehat{\balpha}^{\text{base}}_{0} - \balpha_{0} \right \|_2 \leq \frac{6\sqrt{s_0} \lambda_0}{2c_{0,0} - 3\sqrt{s_{0}} \lambda_0}, 
 \end{align} 
\text{ and }
\begin{align}\label{eq:saebb2}
 \left \| \widehat{\balpha}^{\text{base}}_{0} - \balpha_{0} \right \|_{\infty} \leq \frac{6\sqrt{s_0} \lambda_0}{2c_{0,0} - 3\sqrt{s_{0}} \lambda_0}.
 \end{align}
\end{subequations} 
\end{proposition}

Henceforth, let $\gamma_0 = 6\sqrt{s_0} \lambda_0/(2c_{0,0} - 3\sqrt{s_{0}} \lambda_0)$. 
\subsection{Properties of the Simple Average Estimator}\label{sec:psae}

Let $\| \bar \bdelta \|_0 = s_{\bar \bdelta}$ and $\| \bar \bdelta \|_1 = h$. Apparently, $h \leq \| \bar \bdelta \|_0 \cdot \| \bar \bdelta \|_{\infty} \leq s_{\bar \bdelta} \| \bar \bdelta \|_{\infty}$. Furthermore, let $\gamma_{\bar \bdelta} =  2\{(1/K) \sum_{i = 1}^K \gamma_{i} + \gamma_0\}$, $ \sigma_1 = (1/K)\sum_{i=1}^K 4 s_{i} \lambda_i/(c_{i,0} - 2 \sqrt{s_{i} \lambda_{i}}) + 4\sqrt{2s_{\bar \bdelta}h\gamma_{\bar\bdelta}}$, and $\sigma_2 = (1/K)\sum_{i=1}^K4\sqrt{s_i} \lambda_i/(c_{i,0} - 2 \sqrt{s_{i}} \lambda_{i}) \ + \sqrt{2h\gamma_{\bar\bdelta}}$. Then, we can obtain the following convergence rates of the SAE. 

\begin{theorem}\label{thm:sae}
Under the conditions of Proposition~\ref{prop:bs-bound} and Assumption~\ref{ass:d}, $\exists N_{i}, i \in [K]$ large enough, so that if $n_i > N_{i}$ jointly hold, then, on the event $\left(\cup_{i \in [K]}\cup_{j \in [d]_+} \cE_{i,j}\right)^c$, we obtain the following: 
\begin{subequations}
\begin{align}\label{eq:saebs0}
 \left \| \widehat{\balpha}^{\text{sae}}_{0} - \balpha_{0} \right \|_2 \leq \frac{2\sigma_2}{c_{0,0} - \sigma_2}
\end{align} 
 \text{ and }
 \begin{align}\label{eq:saebs1}
 \left \| \widehat{\balpha}^{\text{sae}}_{0} - \balpha_{0} \right \|_1 \leq \frac{\sigma_1+\sqrt{s_0} \sigma_2}{c_{0,0} - \sigma_2}.     
 \end{align} 
\end{subequations}
\end{theorem}

\begin{remark}\label{rem:sae}
Let $b^{\text{base}}_{2}$, $b^{\text{base}}_{1}$, $b^{\text{sae}}_{2}$ and $b^{\text{sae}}_{1}$ be the upper bounds in Inequalities~\ref{eq:saebb0},~\ref{eq:saebb1},~\ref{eq:saebs0} and~\ref{eq:saebs1}, respectively. Combining Proposition~\ref{prop:bs-bound} and Theorem~\ref{thm:sae}, we obtain that
for  $n_0$ large enough, $\exists$ $n_i$, $i \in [K]_+$ large enough such that on the event $\left(\cup_{i \in [K]}\cup_{j \in [d]_+} \cE_{i,j}\right)^c$, $b^{\text{sae}}_{2} \asymp \sqrt{\sqrt{s_0}h\lambda_{0}}$ and $b^{\text{sae}}_{1} \asymp 
\sqrt{\sqrt{s_0}s_{\bar \bdelta}h\lambda_{0}}$, while  $b^{\text{base}}_{2} \asymp \sqrt{s_0}\lambda_0$, $b^{\text{base}}_{1} \asymp s_0\lambda_0$.
If $h = o(\sqrt{s_0} \lambda_0)$, especially, if $s_{\bar \bdelta} \| \bar \bdelta \|_{\infty} = o(\sqrt{s_0} \lambda_0)$,  then $b^{\text{sae}}_{2} = o(b^{\text{base}}_{2})$. Furthermore, if $s_{\bar \bdelta} = O(s_0)$ holds, $b^{\text{sae}}_{1} = o(b^{\text{base}}_{1})$.     
The potential improvement in convergence rates demonstrates the advantage of the SAE over the baseline estimator. This gain arises from two sources: first, the positive transfer of knowledge from the source data, which relies on the sparser structure of the contrast; second, the more accurate estimation of score functions in the source domain, which renders the estimation error of $\bar \bdelta$ negligible.
\end{remark}

\subsection{Properties of the Optimal-Convex-Combination Estimator}\label{sec:occe}

For any $i \in [K]_+$, let $\| \bdelta_i \|_0 = s_{\bdelta_i}$ and $\| \bdelta_i \|_1 = h_i$. Then, $h_i \leq \| \bdelta_i \|_0 \cdot \| \bdelta_i \|_{\infty} \leq s_{\bdelta_i} \| \bdelta_i \|_{\infty}$. Moreover, let $\gamma_{\bdelta_i} = 2(\gamma_i + \gamma_0)$, $\sigma_{i,1} = 4 s_{i} \lambda_i/(c_{i,0} - 2 \sqrt{s_{i}} \lambda_{i}) + 4\sqrt{2s_{\bdelta_i}h_i\gamma_{\bdelta_i}}$ and $\sigma_{i,2} = 4\sqrt{s_i} \lambda_i/(c_{i,0} - 2 \sqrt{s_{i}} \lambda_{i}) + \sqrt{2h_i\gamma_{\bdelta_i}}$.

For the source estimators, we establish the following high-probability error bounds.

\begin{proposition}\label{prop:s-e}
Under the conditions of Lemma~\ref{lem:sbu-g} and Assumptions~\ref{ass:order} to~\ref{ass:mu}, for any $i \in [K]_+$, $\exists N_{i}$ large enough, so that if $n_i > N_{i}$, then, on the event $\left(\cup_{k \in \{0,i\}}\cup_{j \in [d]_+} \cE_{k,j}\right)^c$, we obtain the following: 
\begin{subequations}
\begin{align}
\left \| \widehat{\balpha}_{0,i} - \balpha_{0} \right \|_2 \leq \frac{2\sigma_{i,2}}{c_{i,0} - 2 \sqrt{s_{i}} \lambda_{i}},
\end{align}
and
\begin{align}
 \left \| \widehat{\balpha}_{0,k} - \balpha_{0} \right \|_1 \leq  \frac{\sigma_{i,1}+\sqrt{s_0} \sigma_{i,2}}{c_{i,0} - 2 \sqrt{s_{i}} \lambda_{i}}.    
\end{align}
\end{subequations}
\end{proposition}

For any estimator $\widehat{\balpha}_0$ of $\balpha_0$, define its empirical MSE and population MSE respectively as 
\begin{align*}
\widehat{\text{MSE}}(\widehat{\balpha}_0) = \left \| \widehat{\balpha}_0^{\text{base}} -\widehat{\balpha}_0 \right \|^2_{2}  
\text{ and } 
\text{MSE}(\widehat{\balpha}_0) = \left \| \widehat{\balpha}_0 - \balpha_0 \right \|^2_{2}.
\end{align*}

Now consider a non-stochastic dictionary of $\ell$ estimates of $\balpha_0$, denoted by $
\cH = \{\widehat{\balpha}_{0,1}, \ldots, \widehat{\balpha}_{0,\ell} \}$, and let $r^{*}_{\cH} = \sup_{i \in [\ell]_+} \| \widehat{\balpha}_{0,i} - \balpha_0 \|_{2}$.    

Proposition~\ref{prop:data-driven} provides a high‑probability upper bound for the MSE of the optimal convex combination estimator of the dictionary .

\begin{proposition}\label{prop:data-driven}
Under the conditions of Proposition~\ref{prop:bs-bound} and Assumption~\ref{ass:mu}, further assume that for the non-stochastic dictionary $\cH$, $r^{*}_{\cH} < 1$, for large enough $n_0$, on the event $\left(\cup_{j \in [d]_+} \cE_{0,j}\right)^c$,
\begin{equation}\label{eq:mse}
\begin{aligned}
 \text{MSE}(\widehat{\balpha}_{0,\widetilde{\bgamma}}) 
 \leq \inf_{\bgamma \in \Delta^{\ell}} \text{MSE}(\widehat{\balpha}_{0,\bgamma}) +\frac{24\sqrt{s_0} r^{*}_{\cH} \lambda_0}{(1 - r^{*}_{\cH})(2c_{0,0} - 3\sqrt{s_{0}} \lambda_0)}.     
\end{aligned}
\end{equation}         
\end{proposition}

We now derive explicit convergence guarantee for the optimal convex combination estimator $\widehat{\balpha}_{0,\widetilde{\bgamma}}$. This result follows from Proposition~\ref{prop:data-driven} and is stated below.

\begin{theorem}
\label{thm:oca}
Under the conditions of Proposition~\ref{prop:data-driven} and Assumption~\ref{ass:order}, for any $n_0 \in \NN_+$, $\exists N_k(c_{0,0},n_0) $ for $k \in [K]_+$ and some constants $c_{1,6}, c_{2,6} > 0$ w.r.t $n_i$, so that for any $n_k > N_k(c_{0,0},n_0)$, with probability at least $1 - \sum_{i=0}^{K}d/n_{i}^2$, the estimator $\widehat{\balpha}_{0,\widetilde{\bgamma}}$ obtained from Algorithm~\ref{alg:gamma_combination} satisfies 
\begin{align}\label{eq:occe}
\text{MSE}(\widehat{\balpha}_{0,\widetilde{\bgamma}}) \leq \inf_{i \in [K]_+} \frac{4\sigma^2_{i,2}}{(c_{i,0} - 2 \sqrt{s_{i}} \lambda_{i})^2} + \frac{24\sqrt{s_0}  \lambda_0}{(1 - c)(2c_{0,0} - 3\sqrt{s_{0}} \lambda_0)}\sup_{i\in [K]_+} \frac{2\sigma_{i,2}}{c_{i,0} - 2 \sqrt{s_{i}} \lambda_{i}}.
\end{align}
\end{theorem}

\begin{remark}
Let $b^{\text{base}}_{2}$, and $b^{\text{occ}}_{2}$ be the upper bounds in Inequalities~\ref{eq:saebb0} and~\ref{eq:occe}, respectively. Combining Proposition~\ref{prop:bs-bound} and Theorem~\ref{thm:sae}, we obtain that
for $n_0$  large enough, and any $i \in [K]_+$, $\exists$ $N_i > 0$, such that for any $n_i > N_i$, on the event $\left(\cup_{k \in \{ 
 0,i\}}\cup_{j \in [d]_+} \cE_{k,j}\right)^c$, $b^{\text{occ}}_{2}  \asymp \{\sqrt{s_0}h_i\lambda_{0} + (\sqrt{s_i} \lambda_i +  \sqrt{\sqrt{s_0}h_i\lambda_{0}})\sqrt{s_0}\lambda_0\}$, while $(b^{\text{base}}_{2})^2 \asymp s_0\lambda^2_0$. Similar to the discussion in Remark~\ref{rem:sae}, by Definition~\ref{def:inf}, if all sources are informative, $b^{\text{occ}}_{2} = o\{(b^{\text{base}}_{2})^2\}$. 
\end{remark}

Comparing the error bounds of SAE and OCCE in Inequalities~\eqref{eq:saebs0} and~\eqref{eq:occe}, respectively, we obtain the following corollary regarding their relationship.

\begin{corollary}\label{cor:com-bd}
Let $\kappa_{\text{SAE}}$ denote the square of the error bound in Inequality ~\eqref{eq:saebs0} and $\kappa_{\text{OCCE}}$ denote the error bound in Inequality~\eqref{eq:occe}.
Under the conditions of Theorem~\ref{thm:sae} and Theorem~\ref{thm:oca}, we have the following relationship between $\kappa_{\text{SAE}}$ and $\kappa_{\text{OCCE}}$:
\begin{itemize}
    \item [1. ]If $\left\{s_i^{1/2}\lambda_i+s_0^{1/4}(h_i\lambda_0)^{1/2}\right\} = o(h)$, $\kappa_{\text{OCCE}} = o(\kappa_{\text{SAE}})$.
    \item [2. ] If $\left\{ s_i^{1/2}\lambda_i+s_0^{1/4}(h_i\lambda_0)^{1/2}\right\} \asymp h$, $\kappa_{\text{OCCE}} \asymp \kappa_{\text{SAE}}$.
    \item [3. ] If $h = o\left\{ s_i^{1/2}\lambda_i+s_0^{1/4}(h_i\lambda_0)^{1/2}\right\}$, $ \kappa_{\text{SAE}} =  o(\kappa_{\text{OCCE}})$.
\end{itemize}
\end{corollary}

\begin{remark}
The result in Corollary~\ref{cor:com-bd} aligns with intuition: if the source indices are complementary such that $\bar{\bdelta}$ remains sufficiently sparse, the SAE can accelerate convergence by pooling more samples across similar sources. Conversely, if the source indices are so heterogeneous that the sparsity of $\bar{\bdelta}$ is violated, then, under the assumption that all sources are informative, the combined sources, even if individually informative, may jointly perform worse than a single worst source. In the latter case, the OCCE is preferable.
\end{remark}
%In Section~\ref{sec:e-e}, we illustrate a scenario where OCCE outperforms SAE, while a case that favors SAE is presented in Section~\ref{sec:rda}.

\subsection{Properties of Informative-Source Selection Procedure}\label{sec:issp}

To distinguish between non-informative and informative sources, some extra assumptions are necessary.

\begin{assumption}\label{ass:d-m} 
For large enough $n_0$,
$\exists \delta(n_0) > 0$, such that for any informative source
$i$, and 
any $j \in \{ j \in [d]_+ | [\bdelta_i]_j  \neq 0 \}$, $|[\bdelta_i]_j | < \{1+\delta(n_0)\} \gamma_0$, while for non-informative source $i'$, exists $j' \in \{ j \in [d]_+ | [\bdelta_{i'}]_j  \neq 0 \}$, such that $\gamma_0 = o(|[\bdelta_{i'}]_{j'}|)$. 
\end{assumption}

To distinguish informative from non-informative sources, Assumption~\ref{ass:d-m} requires that for the former, all non-zero entries of their contrast (relative to the source domain) be sufficiently small, while for the latter, at least one such entry exceed a detectable threshold. Under Assumption~\ref{ass:d-m}, we establish the following ``consistency'' result for the informative-source selection procedure.

\begin{proposition}\label{prop:s}
Suppose Assumptions~\ref{ass:sf} to~\ref{ass:d-m} hold, for $\forall \xi > 0$, $\exists N_0(\xi) \in \NN_+$  such that for $n_0 > N_0(\xi)$, $\exists N(\xi, n_0) \in \NN_+$, if $\min_{i \in [K]_+} n_i > N(\xi,n_0)$, then, if $k$ equals to the true number of informative sources, with probability at least $1 - \xi$, sources in $\sI$ are all informative while sources not in are all non-informative.  
     
\end{proposition}

\begin{remark}
Proposition~\ref{prop:s} guarantees that, as $n_0$ and $n_i$ tend to infinity in a sequential manner, the selection procedure is ``consistent'' with high probability. However, it requires strong prior knowledge to know the true number of informative sources or users need to take conservative strategies, i.e, select a small number of source domains to avoid negative transfer.  
\end{remark}

\subsection{Error Decomposition of the Composed Estimator}\label{sec:ed}

This subsection presents an error decomposition analysis for the composed estimator. We first state the following regularity conditions required for the analysis.

\begin{assumption}\label{ass:bx}
For $\bx^{(0)} = ([\bx^{(0)}]_1, \ldots, [\bx^{(0)}]_d)^\T$, $[\bx^{(0)}]_1, \ldots, [\bx^{(0)}]_d$ are i.i.d sub-Gaussian with mean 0.  
\end{assumption}

\begin{assumption}\label{ass:be}
$\epsilon_0$ follows a symmetric sub-Gaussian distribution with mean 0 and variance $\sigma^2$. 
\end{assumption}

\begin{assumption}\label{ass:db}
$f_0$ is $L_0$-Lipschitz. 
\end{assumption}

\begin{assumption}\label{ass:f}
The function class $\widehat{f}_{(\cdot)}$ is defined on a compact parameter space $\bTheta \subset [-B, B]^{\otimes (3w+1)}$, and the activation function $\phi$ is ReLU. 
\end{assumption}

Assumption~\ref{ass:bx} is commonly used in both theoretical analysis and simulations in literature on Stein's methods~\citep{2017zhuoran,balasubramanian2018,na2019high}. 
Assumption~\ref{ass:be} is similarly mild and holds for standard zero-centered error distributions such as Gaussian and symmetric uniform.
Assumption~\ref{ass:db} ensures boundedness of the gradient of $f_0$, a condition that could be relaxed to hold locally on the input domain. Finally, Assumption~\ref{ass:f} specifies a widely adopted single-layer MLP architecture. The parameter $B$, which bounds the parameters, is not fixed but allowed to grow with the network width $w$. Moreover, for the generalization error to vanish, $B$ must indeed increase with $w$, a condition formally established in Proposition~\ref{prop:tt} below. In practice, such boundedness is often enforced via parameter clipping. Under the stated assumptions, we derive the following upper bound for the expected mean squared error.

\begin{proposition}\label{prop:tt}
Under the conditions of Theorem~\ref{thm:oca} and Assumptions~\ref{ass:bx} to~\ref{ass:f}, for an estimator $\widehat{\balpha}_0$ of $\balpha_0$ with $\|\widehat{\balpha}_0\|_2=1$, the following upper bound on the expected mean squared error holds:
\begin{align*}
& \EE[\{ f_{0}(\balpha_0^{\T} \bx^{(0)}) - \widehat{f}_{\widehat{\btheta}_{\text{e}}}(\widehat{\balpha}_{0}^{\T}\bx^{(0)}) \}^2]\\
= & O(d  \|\balpha_0 - \widehat{\balpha}_0\|_2^2)+   O\left(\frac{w^{5/2} B^4\ln{n}}{\sqrt{n}}\right) + O\left\{ \frac{B^2}{w^2} + B^2 \exp{ \left(\frac{-B^2}{C}\right)}  \right\} +\sigma^2,    
\end{align*} 
for some constant $C > 0$ depending on $\bx^{(0)}$ and $f_0$.
\end{proposition}

\begin{remark}
The upper bound in Proposition~\ref{prop:tt} consists of four components.

The first term arises from the error in estimating the index $\balpha_0$ by $\widehat{\balpha}_0$. It scales linearly with the dimension dd due to the i.i.d. covariate assumption; this dependence can be reduced by imposing sparsity or correlation structures.

The second and third terms together represent the generalization error of the neural network approximating 
$f_0$, evaluated under the data distribution induced by $\widehat{\balpha}_0$, i.e., on $\{\widehat{\balpha}_0^{\T}\bx,\, f_0(\widehat{\balpha}_0^{\T}\bx)+\epsilon\}$.

The second term arises from fitting the network to $n$ samples. Its order is $O\left(w^{5/2} B^4\ln{n}/\sqrt{n}\right)$, showing high-order dependence on width $w$ and parameter bound $B$. This stems from the squared loss analysis: the Lipschitz constant of the loss is proportional to that of the ReLU network class, which scales as $wB^2$; squaring it gives $B^4$and the higher width dependence.

The third term reflects the ability of a finite-width ReLU network to approximate $f_0$ over the relevant domain.  A trade-off arises: for the approximation error to vanish, $B$ must tend to infinity, but slower than $w$. This is because without a bounded input assumption, finite-width ReLU networks grow linearly outside a compact set; controlling the expected MSE requires expanding the effective domain, forcing $B \rightarrow \infty$. In practice, inputs can be bounded (e.g., via normalization), so $B$ need not diverge and can even decrease with $w$, alleviating the high-order dependence~\citep{bartlett2017spectrally}. However, this practical observation is beyond our core theoretical focus; we present the general bounds without assuming input boundedness.

Finally, the fourth term stems from the irreducible error of the random noise
$\epsilon$.
\end{remark}

\section{Simulation Studies}\label{sec:ss}
The simulation study comprises two main components. First, we evaluate the performances of the SAE and OCCE in recovering the index $\balpha_0$, as well as the performance of the composite estimators in estimating the function $f_0\{\balpha_0^\T (\cdot)\}$. Second, we demonstrate the efficacy of the proposed informative-source selection algorithm. Throughout the experiments, the covariates $\bx^{(i)}$, for $i \in [K]$ are generated from $\cN(\boldsymbol{0},\bSigma)$, where $\bSigma_{i,j} = \rho^{|i-j|}$, for $i,j \in [d]_+$. We set an equal sample size across all sources, i.e., $n_{i} = n$ for each $i \in [K]_{+}$. In the study, we use a deep neural network with denoise score matching as the score function.  
All details of the design of the experiments and extensions are deferred to Section $\mathrm{II.3}$ of the supplement.  

\subsection{Performances of Proposed Estimators}\label{sec:exp-est}

In this part, the observations are generated in the following procedure. 

\subsubsection{the Data Generating Process}
\begin{itemize}
    \item[1. ] Generate $\balpha_0$.
    \begin{itemize}
        \item[$\mathrm(i)$] Generate $s_0$ i.i.d samples from $\P_{p}$%\footnote{In this study, we choose $\P_p$ to be $\text{Uniform}([-1, -0.9] \cup [0.9, 1])$.}
        , denoted as $\boldsymbol{a} = ([\ba]_1, \ldots, [\ba]_{s_0})^\T$.
        \item[$\mathrm(ii)$] Let $[\balpha^{\ddag}_0]_{1:s_0} = \boldsymbol{a}$ and $[\balpha^{\ddag}_0]_{(s_0+1):d} = \boldsymbol{0}$.
        \item[$\mathrm(iii)$] Let $\balpha_0 = \balpha^{\ddag}_0/\| \balpha^{\ddag}_0 \|_2$.
    \end{itemize}
    \item [2. ] For $i \in [K]_+$, generate $\balpha_i$.
    \begin{itemize}
        \item [$\mathrm(i)$] For any $i \in [K]_+$, sample $\cI_{i_o} = \{p_{i,o_1},\ldots,p_{i,o_{d_i}}\}$ uniformly from $[s_0]_+$ without replacement.
        \item [$\mathrm(ii)$] Sample $\cI_{i_t} = \{p_{i,t_1}, \ldots, p_{i,t_{d_i}}\}$ uniformly from $[d]_+\setminus [s_0]_+$ without replacement.
        \item [$\mathrm(iii)$] 
        Let $[\balpha^{\ddag}_i]_{\cI_{i_t}} = [\balpha^{\ddag}_0]_{\cI_{i_o}}$ and $[\balpha^{\ddag}_0]_{\cI_{i_o}} = \zeta_i%\footnote{We let $\rho_i = 0.1$ for any $\i \in [K]_+$.} 
        \cdot[\balpha^{\ddag}_i]_{\cI_{i_o}}$.
        \item[$\mathrm(iv)$] Let $\balpha_i = \balpha^{\ddag}_i/\| \balpha^{\ddag}_i \|_2$.
    \end{itemize}
    
    \item [3. ] For $i \in [K]_+$, generate $n_i$ i.i.d. training samples from the $i$-th source and $n_0$ i.i.d. training samples as well as the testings set $\cS_{\text{test}}$ consisting of $n_{test}$ i.i.d. samples from the target based on Model~\eqref{eq:sim}.  
\end{itemize}

\subsubsection{Competing Methods}\label{sec:cm}
We compare the proposed methods with several competing approaches: a baseline estimator, a target‑domain‑only neural network (NN) method, and a target‑domain‑only lasso estimator. Because our setup assumes no access to source data, very few existing transfer learning methods in the literature can be fairly compared. Nevertheless, we include the transfer learning method $\mathcal{A}$-Trans-GLM introduced by \citet{Tian02102023}, which reduces to a linear method under our setting.

We aim to demonstrate three key points. First, even when only estimated indices are accessible, combining a larger and more diverse set of informative indices yields more accurate estimation of the target index, approaching the performance of the pooling estimators in \citet{Tian02102023}. Second, although individual indices can be helpful, the performance of pooling estimators may deteriorate as more informative sources are added, due to the influence of varying link functions. Third, guidance from pre‑trained indices helps mitigate the overfitting problem of MLP when target samples are scarce, enabling it to generalize better than linear methods ($\mathcal{A}$-Trans-GLM).
%Strictly speaking, a direct comparison with $\cA$-Trans-GLM is not entirely fair, as it assumes full access to source data, which can be pooled with target data to improve estimation. In contrast, our setup only allows summarized statistics i.e., $\widehat{\balpha}_i$, from source domains. Nevertheless, we include this comparison to highlight a key limitation of linear methods, as exemplified by $\cA$-Trans-GLM. Although they serve as strong baselines, especially when source data are available and the underlying models are sufficiently similar in a linear sense, they fail to effectively exploit the useful information hidden under a ``weak non-linear coat'', by which we mean the link functions defined in Assumption~\ref{ass:lf}. In contrast, the proposed methods are specifically designed to capture and leverage such information, enabling them to perform better than linear methods in non-linear cases, even when no source data are available.

For the NN method, we first obtain $\balpha^{\dagger}_{0, n}$ and $\widehat{\btheta}_{\text{e}}(\balpha^{\dagger}_{0, n})$ by solving
\begin{align*}
\balpha^{\dagger}_{0, n}, \widehat{\btheta}_{\text{e}}(\balpha^{\dagger}_{0, n}) = \argmin_{\balpha \in \RR^d, \btheta \in \bTheta} \left (\te \left[ { y - \widehat{f}_{\btheta}(\balpha^{\T}\bx) }^2 \right] + \lambda \|\balpha \|_1 \right),    
\end{align*}
where $\lambda$ is a tuning parameter. The target index is estimated by $\widehat{\balpha}_{0,n} = \balpha^{\dagger}_{0, n} / \| \balpha^{\dagger}_{0, n} \|_2$, and the composed estimator is itself, i.e., $g_{\widehat{\alpha}_{0,n}} \coloneqq \widehat{f}_{\widehat{\btheta}_{\text{e}}(\balpha^{\dagger}_{0, n})}\left\{(\balpha^{\dagger}_{0, n})^{\T} (\cdot)\right\}$.

For the target-domain-only lasso estimator, let $\balpha^\dagger_{0,\ell} = \{\alpha^{\dagger}_b, (\balpha^\dagger_w)^\T\}^\T$ denote the lasso estimator on the target domain. Then, $\balpha_0$ is estimated by $\widehat{\balpha}_{0,\ell} = \balpha^\dagger_w / \|\balpha^\dagger_w\|_2$ and $f_0\{\balpha_0^{\T}(\cdot)\}$ is estimated by the linear model, i.e., $g_{\widehat{\balpha}_{0,\ell}} \coloneqq \alpha^{\dagger}_b + (\balpha^\dagger_w)^\T (\cdot)$. For $\cA$-Trans-GLM, the index estimator $\widehat{\balpha}_{0,t\ell}$ and corresponding composed estimator $g_{\widehat{\balpha}_{0,t\ell}}$ are obtained in the same way.

\subsubsection{Comparison}\label{sec:e-e}

In this part, we first evaluate the performance of proposed methods in estimating the target index $\balpha_0$. For an estimator $\widehat{\balpha}_0$, we provide $\| \widehat{\balpha}_0 - \balpha_0\|_1$, $\| \widehat{\balpha}_0 - \balpha_0 \|_2$ and the angle between $\widehat{\balpha}_0 $ and $\balpha_0$ denoted by $\mathcal{\epsilon}_{L^1}$, $\mathcal{\epsilon}_{L^2}$ and $\mathcal{\epsilon}_{A}$, respectively. Due to the inherent unidentifiability of the sign, we align the estimator’s orientation with that of the true $\balpha_0$ before computing these distances. Then, the performances of the composite estimators are evaluated by the out-of-sample $R$-squared on the test set, which is defined as
\begin{align*}
R^{2}_{\text{oos}} = \frac{1 - \te_{\cS_{\text{test}}}\left\{g_{\widehat{\balpha}_0}(\bx^{(0)}) - y^{(0)}\right \}^{2}}{ \te_{\cS_{\text{test}}}\left \{  y^{(0)} -  \te_{\cS_{\text{test}}}  \left ( y^{(0)} \right ) \right\}^{2}}.  
\end{align*}

In this study, we consider $K = 1, 4, 7$ and three different combinations of link functions: 
\begin{enumerate}
    \item [] \textbf{Case 1.} All domains share the same link function $f(x) = erf(x)$. 
    \item [] \textbf{Case 2.} For $i \in [K]$, the link function of the $i$-th domain is $f_i(x) = tanh(x - 0.25i)$.
    \item [] \textbf{Case 3.} In this case, we explore link functions in a more diversified way. Specifically, we let
    $f_0(x) = tanh(x)$, $f_1(x) = erf(x -0.25)$, $f_2(x) = sin(x-0.25)$, $f_3(x) = sigmoid(x-0.25)$, $f_4(x) = tanh(x - 0.25)$, $f_5(x) = erf(x + 0.5)$, $f_6(x) = sin(x + 0.5)$, and $f_7(x) = sigmoid(x + 0.5)$.   
\end{enumerate}

\begin{table}[htbp]
\centering
\scriptsize
\caption{Comparison of performances of competing methods in Case 1 over 100 repetitions.}
\label{tb:c1}
\begin{tabular}{lccccc}
\toprule
\multicolumn{6}{c}{\(K=1\)} \\
\cmidrule(lr){1-6}
Index Estimators & $\mathcal{\epsilon}_{L^2}$ & $\mathcal{\epsilon}_{L^1}$ & $\mathcal{\epsilon}_{A}$  & Composed Estimators & $R^2_{\text{oos}}$ \\
\midrule
$\widehat{\balpha}_{0,\ell}$     &  0.2803±0.0286 &  1.8902±0.2057  &  16.1±1.7 & $g_{\widehat{\balpha}_{0,\ell}}$ &  0.5581±0.0167 \\
$\widehat{\balpha}_{0,t\ell}$   & \textbf{0.1473±0.0256}  & \textbf{0.7867±0.1714}  &  \textbf{8.5±1.5}  & $g_{\widehat{\balpha}_{0,t\ell}}$ &0.5790±0.0138 \\
$\widehat{\balpha}_{0,n}$&  0.3253±0.0282  & 2.5983±0.2245 & 8.7±1.6  & $g_{\widehat{\balpha}_{0,n}}$ &  \textbf{0.5871±0.0233}
  \\
$\widehat{\balpha}^{\text{base}}_{0}$ & 1.0062±0.1115 & 7.9553±0.8838  & 60.6±7.5 & $g_{\widehat{\balpha}^{\text{base}}_{0}}$ &  0.2086±0.0788  \\
$\widehat{\balpha}^{\text{sae}}_{0}$  & 0.3781±0.0479  & 2.8427±0.5670  & 21.8±2.8   & $g_{\widehat{\balpha}^{\text{sae}}_{0}}$ &  0.5518±0.0304 \\
$\widehat{\balpha}^{\text{occ}}_{0}$  & 0.3781±0.0479  & 2.8427±0.5670  & 21.8±2.8 & $g_{\widehat{\balpha}^{\text{occ}}_{0}}$ & 0.5505±0.0303 \\
\midrule
\multicolumn{6}{c}{\(K=4\)} \\
\cmidrule(lr){1-6}
Index Estimators & $\mathcal{\epsilon}_{L^2}$ & $\mathcal{\epsilon}_{L^1}$ & $\mathcal{\epsilon}_{A}$  & Composed Estimators & $R^2_{\text{oos}}$ \\
\midrule
$\widehat{\balpha}_{0,\ell}$     & 0.2766±0.0347 & 1.8649±0.2447 & 15.9±2.0  & $g_{\widehat{\balpha}_{0,\ell}}$ & 0.5596±0.0162 \\
$\widehat{\balpha}_{0,t\ell}$   & \textbf{0.0987±0.0216 } &  \textbf{0.5270±0.1533} &  \textbf{5.7±1.2}  & $g_{\widehat{\balpha}_{0,t\ell}}$ & 0.5854±0.0140  \\
$\widehat{\balpha}_{0,n}$& 0.3210±0.0271 & 2.5752±0.2392  & 18.5±1.6 & $g_{\widehat{\balpha}_{0,n}}$ & 0.5885±0.0233  \\
$\widehat{\balpha}^{\text{base}}_{0}$ &  0.9957±0.1191 & 7.8634±0.9288 &   59.9±8.1 & $g_{\widehat{\balpha}^{\text{base}}_{0}}$ & 0.2101±0.0787 \\
$\widehat{\balpha}^{\text{sae}}_{0}$  & 0.2113±0.0375  &  1.6572±0.2928 &  12.1±2.2  & $g_{\widehat{\balpha}^{\text{sae}}_{0}}$ &  \textbf{ 0.6134±0.0217 } \\
$\widehat{\balpha}^{\text{occ}}_{0}$  & 0.2632±0.1213  & 2.0421±0.9523   & 15.2±7.4   & $g_{\widehat{\balpha}^{\text{occ}}_{0}}$ & 0.5930±0.0691  \\
\midrule
\multicolumn{6}{c}{\(K=7\)} \\
\cmidrule(lr){1-6}
Index Estimators & $\mathcal{\epsilon}_{L^2}$ & $\mathcal{\epsilon}_{L^1}$ & $\mathcal{\epsilon}_{A}$  & Composed Estimators & $R^2_{\text{oos}}$  \\
\midrule
$\widehat{\balpha}_{0,\ell}$ & 0.2744±0.0299    &  1.8370±0.2186  &  15.8±1.7  & $g_{\widehat{\balpha}_{0,\ell}}$ &  0.5612±0.0138  \\
$\widehat{\balpha}_{0,t\ell}$   & \textbf{ 0.0770±0.0219}  & \textbf{ 0.4309±0.1432}  & \textbf{4.4±1.3 }  & $g_{\widehat{\balpha}_{0,t\ell}}$ & 0.5890±0.0121  
\\
$\widehat{\balpha}_{0,n}$&  0.3216±0.0338 &  2.5619±0.2894  & 18.5±2.0  & $g_{\widehat{\balpha}_{0,n}}$ &  0.5896±0.0200 \\
$\widehat{\balpha}^{\text{base}}_{0}$ & 0.9891±0.1180   & 7.8458±0.9442   &  59.4±8.0  & $g_{\widehat{\balpha}^{\text{base}}_{0}}$ & 0.2141±0.0796 \\
$\widehat{\balpha}^{\text{sae}}_{0}$ & 0.1632±0.0169 & 1.2877±0.1417   & 9.4±1.0   & $g_{\widehat{\balpha}^{\text{sae}}_{0}}$ & \textbf{ 0.6275±0.0185} \\
$\widehat{\balpha}^{\text{occ}}_{0}$  & 0.2386±0.1152  &  1.8590±0.9173  &  13.8±7.2   & $g_{\widehat{\balpha}^{\text{occ}}_{0}}$ & 0.6045±0.0634  \\
\bottomrule
\end{tabular}
\par\footnotesize Note: Data are presented as mean±std.
\end{table}

\begin{table}[htbp]
\centering
\scriptsize
\caption{Comparison of performances of competing methods in Case 2 over 100 repetitions.}\label{tb:c2}
\begin{tabular}{lccccc}
\toprule
\multicolumn{6}{c}{\(K=1\)} \\
\cmidrule(lr){1-6}
Index Estimators & $\mathcal{\epsilon}_{L^2}$ & $\mathcal{\epsilon}_{L^1}$ & $\mathcal{\epsilon}_{A}$ & Composed Estimators & $R^2_{\text{oos}}$ \\
\midrule
$\widehat{\balpha}_{0,\ell}$  & 0.2951±0.0289 & 1.9901±0.2154 & 17.0±1.7  & $g_{\widehat{\balpha}_{0,\ell}}$ &  0.5320±0.0188 \\
$\widehat{\balpha}_{0,t\ell}$   & \textbf{0.1610±0.0253 } & \textbf{0.8547±0.1621 } & \textbf{9.2±1.5 }  & $g_{\widehat{\balpha}_{0,t\ell}}$ & \textbf{0.5523±0.0157 } \\
$\widehat{\balpha}_{0,n}$& 0.3566±0.0305& 2.8507±0.2432   &  20.5±1.8  & $g_{\widehat{\balpha}_{0,n}}$ &0.5424±0.0264 \\
$\widehat{\balpha}^{\text{base}}_{0}$ & 1.0091±0.1100  & 7.9764±0.8785  & 60.7±7.4 & $g_{\widehat{\balpha}^{\text{base}}_{0}}$ &  0.1986±0.0739 \\
$\widehat{\balpha}^{\text{sae}}_{0}$  & 0.3841±0.0465& 2.8979±0.5398   & 22.1±2.7 & $g_{\widehat{\balpha}^{\text{sae}}_{0}}$ & 0.5199±0.0298 \\
$\widehat{\balpha}^{\text{occ}}_{0}$  &  0.3841±0.0465  & 2.8979±0.5398 & 22.1±2.7  & $g_{\widehat{\balpha}^{\text{occ}}_{0}}$ &  0.5191±0.0298 \\
\midrule
\multicolumn{6}{c}{\(K=4\)} \\
\cmidrule(lr){1-6}
Index Estimators & $\mathcal{\epsilon}_{L^2}$ & $\mathcal{\epsilon}_{L^1}$ & $\mathcal{\epsilon}_{A}$ & Composed Estimators & $R^2_{\text{oos}}$ \\
\midrule
$\widehat{\balpha}_{0,\ell}$  & 0.2917±0.0356  & 1.9667±0.2571 & 16.8±2.1 & $g_{\widehat{\balpha}_{0,\ell}}$ & 0.5332±0.0180 \\
$\widehat{\balpha}_{0,t\ell}$  & \textbf{0.1265±0.0336} & \textbf{0.7095±0.2499}  &\textbf{7.3±1.9}  & $g_{\widehat{\balpha}_{0,t\ell}}$ & 0.5521±0.0155 \\
$\widehat{\balpha}_{0,n}$& 0.3518±0.0307  & 2.8203±0.2697  & 20.3±1.8   & $g_{\widehat{\balpha}_{0,n}}$ & 0.5443±0.0257  \\
$\widehat{\balpha}^{\text{base}}_{0}$ & 0.9980±0.1183  & 7.8802±0.9239  & 60.0±8.0  & $g_{\widehat{\balpha}^{\text{base}}_{0}}$ & 0.1991±0.0748 \\
$\widehat{\balpha}^{\text{sae}}_{0}$  & 0.2206±0.0409 & 1.7380±0.3232  & 12.7±2.4  & $g_{\widehat{\balpha}^{\text{sae}}_{0}}$ & \textbf{ 0.5747±0.0255 } \\
$\widehat{\balpha}^{\text{occ}}_{0}$  & 0.2727±0.1285  & 2.1194±1.0118  &  15.7±7.8  & $g_{\widehat{\balpha}^{\text{occ}}_{0}}$ &  0.5549±0.0691 \\
\midrule
\multicolumn{6}{c}{\(K=7\)} \\
\cmidrule(lr){1-6}
Index Estimators & $\mathcal{\epsilon}_{L^2}$ & $\mathcal{\epsilon}_{L^1}$ & $\mathcal{\epsilon}_{A}$ & Composed Estimators & $R^2_{\text{oos}}$  \\
\midrule
$\widehat{\balpha}_{0,\ell}$  & 0.2896±0.0304 &1.9392±0.2248 &  16.7±1.8 & $g_{\widehat{\balpha}_{0,\ell}}$ & 0.5347±0.0161   \\
$\widehat{\balpha}_{0,t\ell}$ & 0.2084±0.0322  & \textbf{1.2918±0.2830} & 12.0±1.9  & $g_{\widehat{\balpha}_{0,t\ell}}$ & 0.5429±0.0170 
\\
$\widehat{\balpha}_{0,n}$& 0.3536±0.0381 & 2.8165±0.3171 & 20.4±2.2 & $g_{\widehat{\balpha}_{0,n}}$ &  0.5448±0.0232 \\
$\widehat{\balpha}^{\text{base}}_{0}$ & 0.9912±0.1175  & 7.8641±0.9402   & 59.6±7.9  & $g_{\widehat{\balpha}^{\text{base}}_{0}}$ &  0.2025±0.0762 \\
$\widehat{\balpha}^{\text{sae}}_{0}$ & \textbf{0.1803±0.0158}  &  1.4334±0.1252 & \textbf{10.3±0.9} & $g_{\widehat{\balpha}^{\text{sae}}_{0}}$ & \textbf{0.5886±0.0205} \\
$\widehat{\balpha}^{\text{occ}}_{0}$  & 0.2889±0.1174   & 2.2789±0.9348   &  16.7±7.3  & $g_{\widehat{\balpha}^{\text{occ}}_{0}}$ &  0.5551±0.0623  \\
\bottomrule
\end{tabular}
\par\footnotesize Note: Data are presented as mean±std.
\end{table}

\begin{table}[htbp]
\centering
\scriptsize
\caption{Comparison of performances of competing methods in Case 3 over 100 repetitions.}
\label{tb:c3}
\begin{tabular}{lccccc}
\toprule
\multicolumn{6}{c}{\(K=1\)} \\
\cmidrule(lr){1-6}
Index Estimators & $\mathcal{\epsilon}_{L^2}$ & $\mathcal{\epsilon}_{L^1}$ & $\mathcal{\epsilon}_{A}$  & Composed Estimators & $R^2_{\text{oos}}$ \\
\midrule
$\widehat{\balpha}_{0,\ell}$     & 0.2951±0.0289  &  1.9901±0.2154 &  17.0±1.7   & $g_{\widehat{\balpha}_{0,\ell}}$ &  0.5320±0.0188  \\
$\widehat{\balpha}_{0,t\ell}$   &  \textbf{ 0.1504±0.0246} &  \textbf{ 0.7875±0.1389}  &  \textbf{8.6±1.4 }   & $g_{\widehat{\balpha}_{0,t\ell}}$ &  \textbf{ 0.5552±0.0153}  \\

$\widehat{\balpha}_{0,n}$& 0.3566±0.0305 & 0.3566±0.0305  &   20.5±1.8   & $g_{\widehat{\balpha}_{0,n}}$ &  0.5424±0.0264  \\
$\widehat{\balpha}^{\text{base}}_{0}$   & 1.0091±0.1100   & 7.9764±0.8785   & 60.7±7.4   & $g_{\widehat{\balpha}^{\text{base}}_{0}}$  &  0.1986±0.0739 \\
$\widehat{\balpha}^{\text{sae}}_{0}$& 0.3797±0.0501 &  2.8383±0.6086  & 21.9±2.9    & $g_{\widehat{\balpha}^{\text{sae}}_{0}}$ & 0.5202±0.0311   \\
$\widehat{\balpha}^{\text{occ}}_{0}$  & 0.3797±0.0501   & 2.8383±0.6086   & 21.9±2.9 &$g_{\widehat{\balpha}^{\text{occ}}_{0}}$ &  0.5202±0.0311 
  \\
\midrule
\multicolumn{6}{c}{\(K=4\)} \\
\cmidrule(lr){1-6}
Index Estimators & $\mathcal{\epsilon}_{L^2}$ & $\mathcal{\epsilon}_{L^1}$ & $\mathcal{\epsilon}_{A}$  & Composed Estimators & $R^2_{\text{oos}}$ \\
\midrule
$\widehat{\balpha}_{0,\ell}$ &  0.2917±0.0356  & 1.9667±0.2571   &  16.8±2.1  & $g_{\widehat{\balpha}_{0,\ell}}$ & 0.5332±0.0180 \\
$\widehat{\balpha}_{0,t\ell}$   &  \textbf{0.1859±0.0420}  &  \textbf{1.1232±0.3406}  &  \textbf{10.7±2.4} & $g_{\widehat{\balpha}_{0,t\ell}}$ &   0.5435±0.0156    \\
$\widehat{\balpha}_{0,n}$ & 0.3524±0.0315  &  2.8257±0.2677 & 20.3±1.8  & $g_{\widehat{\balpha}_{0,n}}$ & 0.5439±0.0260  \\
$\widehat{\balpha}^{\text{base}}_{0}$  & 0.9979±0.1178 & 7.8772±0.9161   & 60.0±8.0  & $g_{\widehat{\balpha}^{\text{base}}_{0}}$ &  0.1994±0.0733  \\
$\widehat{\balpha}^{\text{sae}}_{0}$  & 0.2360±0.0419  &  1.8654±0.3340  &  13.6±2.4 & $g_{\widehat{\balpha}^{\text{sae}}_{0}}$ &  \textbf{0.5717±0.0251}   \\
$\widehat{\balpha}^{\text{occ}}_{0}$  &  0.3037±0.1019  & 2.3357±0.8009   &    17.5±6.1 & $g_{\widehat{\balpha}^{\text{occ}}_{0}}$ &  0.5462±0.0540  \\
\midrule
\multicolumn{6}{c}{\(K=7\)} \\
\cmidrule(lr){1-6}
Index Estimators & $\mathcal{\epsilon}_{L^2}$ & $\mathcal{\epsilon}_{L^1}$ & $\mathcal{\epsilon}_{A}$  & Composed Estimators & $R^2_{\text{oos}}$  \\
\midrule
$\widehat{\balpha}_{0,\ell}$ & 0.2896±0.0304   & 1.9392±0.2248  & 16.7±1.8   & $g_{\widehat{\balpha}_{0,\ell}}$ &  0.5347±0.0161  \\
$\widehat{\balpha}_{0,t\ell}$   &  0.1977±0.0338  & \textbf{1.2096±0.2857}   &  11.3±1.9   & $g_{\widehat{\balpha}_{0,t\ell}}$ & 0.5442±0.0156  
\\
$\widehat{\balpha}_{0,n}$& 0.3536±0.0381 &  2.8165±0.3171   & 20.4±2.2   & $g_{\widehat{\balpha}_{0,n}}$ &  0.5448±0.0232 \\
$\widehat{\balpha}^{\text{base}}_{0}$ &  0.9912±0.1175  &  7.8641±0.9402  & 59.6±7.9   & $g_{\widehat{\balpha}^{\text{base}}_{0}}$ &  0.2025±0.0762 \\
$\widehat{\balpha}^{\text{sae}}_{0}$ & \textbf{0.1830±0.0158}  & 1.4544±0.1294   & \textbf{10.5±0.9}  & $g_{\widehat{\balpha}^{\text{sae}}_{0}}$ & \textbf{0.5879±0.0206}  \\
$\widehat{\balpha}^{\text{occ}}_{0}$  & 0.2867±0.1156   &  2.2457±0.9228   &   16.6±7.3  & $g_{\widehat{\balpha}^{\text{occ}}_{0}}$ &  0.5559±0.0628  \\
\bottomrule
\end{tabular}
\par\footnotesize Note: Data are presented as mean±std.
\end{table}

Tables~\ref{tb:c1} to~\ref{tb:c3} summarize the outcomes of the three cases, respectively. We first analyze the estimates of the target index. For the non‑transfer method, $\widehat{\balpha}_{0,\ell}$ performs best, because the underlying link functions are locally close to linear; hence the Lasso estimator suffers little model misspecification and achieves good accuracy. In contrast, $\widehat{\balpha}^{\text{base}}_{0}$ converges more slowly as a moment estimator, and the neural network is designed for the whole function rather than for the index itself.

As anticipated, only when the link functions are identical across domains does the pooled $\cA$-Trans‑GLM estimator improve consistently as more informative sources are added (Table~\ref{tb:c1}), since this is a trivial case for the method: with identical link functions, adding more informative sources simply increases the total sample size by pooling the source data. In other cases, the performance of $\widehat{\balpha}_{0,t\ell}$ degrades as more sources are added, even though each source is individually informative; this is due to the effects of diverse link functions. That $\widehat{\balpha}_{0,t\ell}$ always achieves the smallest $\epsilon_{L^1}$ is not surprising, as it is the objective it directly minimizes.

In contrast, the proposed $\widehat{\balpha}^{\text{sae}}_{0}$ becomes increasingly accurate as more informative sources are added, approaching and even surpassing the best performance of $\widehat{\balpha}_{0,t\ell}$ when link functions vary across domains, even without access to the source data. The proposed $\widehat{\balpha}^{\text{sae}}_{0}$ acts as a safeguard, performing no worse than the worst informative source individually. However, it relies more heavily on the accuracy of the base estimator, which determines the optimal weights. 

The benefit of the nonlinear link functions is significant, and it is also clear that, with guidance from an accurate estimate of the index, the composed estimator can generalize better.

\subsection{the Effectiveness of the Source Selection Algorithm}\label{sec:exp-ssa}

In this section, we assess the performance of the proposed source selection algorithm. After obtaining the target index $\balpha_0$, we create non-informative source indices by a spike perturbation mechanism:

\begin{enumerate}
    \item [1. ] Randomly select $n_s$ dimensions (without replacement) from all $d$ dimensions with equal probabilities;
    \item [2. ] For each selected dimension, add a large spike of magnitude $m_s$, whose sign matches the sign of the corresponding component of  $\balpha_0$;
    \item [3. ] Normalize the vector to unit length.    
\end{enumerate}

We consider three different combinations of link functions. In the first case, all domains share the same link function $f(x) = tanh(x)$. To test the robustness of the proposed method, in the other two cases we fix $f_0(x) = tanh(x)$, and shift the link functions for the source domains to the right by 0.25 and 0.5 units, respectively, i.e., we set $f(x) = tanh(x - 0.25)$ and $f(x) = tanh(x - 0.5)$.

Two informative and two non‑informative sources are considered. After the indices and link functions are determined, observations from informative and non‑informative sources can be generated from Model~\eqref{eq:sim}.

We compare our source‑selection procedure with the method of \citet{Tian02102023}, which provides a built‑in mechanism (Trans-GLM) that automatically decides which source domains should be transferred. For each source domain, the algorithm computes a cross‑validation loss when that source is combined with the target. An internal cross‑validation then determines a threshold; all source domains with a loss below this threshold are selected as beneficial.

The selection mechanism of Trans-GLM provides a natural ordering, whereby a smaller loss indicates a more helpful source. Accordingly, a ranking‑based variant (denoted as Trans-GLM Ranking) is also considered, in which source domains are sorted by their loss in ascending order.

Each source selection method is repeated 100 times. The classification accuracies for informative and non‑informative sources are then calculated. The results are summarised in Table~\ref{tb:ss}. The built‑in thresholding method fails due to non‑linearity. Although the loss is reliable when the source and target models are sufficiently similar, its performance degrades substantially when the models are more dissimilar. Nevertheless, the results presented in Table~\ref{tb:c2} demonstrate that correctly specified informative sources can still improve the target estimator. By contrast, the proposed ordering is considerably more robust to differences in the link functions between the source and target domains, particularly when those differences involve non‑linearity. 

Although comparing the proposed method with the two methods above is inherently unfair, as the latter rely on cross‑validation applied to the source domains, which is inapplicable in our setup, we wish to demonstrate that even with access to source data, the diversity of nonlinear link functions renders linear surrogate losses, friendly to small samples, unreliable, leading automatic threshold selection to fail. In contrast, when the model is correctly specified, a sparsity‑based ranking method is more robust. Provided that sufficient prior knowledge of the number of informative sources is available, or that a conservative strategy is adopted (e.g., selecting only a few informative sources), negative transfer can be avoided with high probability.

\begin{table}[htbp]
\centering
\caption{Classification accuracy of source selection methods over 100 repetitions.}\label{tb:ss}
\begin{tabular}{c S[table-format=2.0, group-digits=false, table-align-text-post=false] 
                S[table-format=2.0, group-digits=false, table-align-text-post=false] 
                S[table-format=2.0, group-digits=false, table-align-text-post=false]}
\toprule
\textbf{Case} & \textbf{Algorithm 1} & \textbf{Trans-GLM Ranking} & \textbf{Trans-GLM} \\
\midrule
1   & 93\% & 100\% & 38\% \\
2   & 88\% & 100\% & 40\% \\
3   & 80\% & 26\% & 2\% \\
\bottomrule
\end{tabular}
\end{table}

\section{Real Data Analysis}\label{sec:rda}

In this section, we apply the proposed methods to the Communities and Crime dataset~\citep{communities_and_crime_183} from the UC Irvine Machine Learning Repository. 

\subsection{Dataset Description}

The Communities and Crime dataset combines socio‑economic data from the 1990 US Census, law enforcement data from the 1990 Law Enforcement Management and Administrative Statistics survey, and crime data from the 1995 FBI Uniform Crime Reporting program. The data contain 2215 observations and 125 predictive features, including community‑level characteristics such as the percentage of urban population, median family income, per‑capita number of police officers, etc. The response variable is the per‑capita violent crime rate (per 100,000 population), which is defined as the sum of reported murders, rapes, robberies, and aggravated assaults. The dataset also includes a state identifier used to define domain partitions for transfer learning experiments.

To mimic sparsity of the indices, after a pre‑processing procedure (deferred to Section $\mathrm{II.4.1}$ of the supplement), we select 50 predictive features: the 20 with the highest mutual information and the 30 with the lowest mutual information, similar to the setup in Section~\ref{sec:ss}. We then obtain two informative source domains and one target domain, with roughly equal sample sizes across the three. On the target domain, 50\% of the observations are randomly partitioned into the training set and the remaining 50\% into the test set. The proposed methods and the competing methods defined in Section~\ref{sec:cm} are applied to the data. The simulation is repeated 500 times.

\subsection{Performances}

The outcomes are summarized in Table~\ref{tab:r2_results}, where we display the 25\%, 50\%, and 75\% quantiles of \(R_{\text{oos}}^{2}\) across competing methods. We observe that although the target-only lasso method and Trans-GLM perform reasonably well, they are still inferior to our proposed methods due to the non-linearity. Furthermore, the number of observations is insufficient to train a randomly initialized neural network, resulting in small or even negative  \(R_{\text{oos}}^{2}\) values, indicating that the estimate is worse than the sample mean. In contrast, with modest guidance on the direction of the index, $g_{\widehat{\balpha}^{\text{base}}_{0}}$ performs substantially better than the previous methods. Moreover, when more precise directions are derived from information transferred from source domains, the estimate can be further improved, not only becoming more precise, but also more robust. The proposed OCCE method works better than SAE for two reasons: first, the base estimator can provide relatively reliable information; second, since due to the generating procedure, the sources tend to be homogeneous.      

\begin{table}[htbp]
\centering
\begin{tabular}{lccc}
\toprule
Estimators & 25\% & 50\% & 75\% \\
\midrule
$g_{\widehat{\balpha}_{0,\ell}}$ & 0.6793&0.7618 & 0.8125\\
$g_{\widehat{\balpha}_{0,t\ell}}$ & 0.7423& 0.8040& 0.8366\\
$g_{\widehat{\balpha}_{0,n}}$ &-0.0676 & 0.2135& 0.4183\\
$g_{\widehat{\balpha}^{\text{base}}_{0}}$ &0.7908 &0.8475 & 0.8828\\
$g_{\widehat{\balpha}^{\text{sae}}_{0}}$ & 0.7965& 0.8525 & 0.8847\\
$g_{\widehat{\balpha}^{\text{occ}}_{0}}$ & \textbf{0.8004}& \textbf{0.8575}& \textbf{0.8863}\\
\bottomrule
\end{tabular}
\caption{The 25\%, 50\%, and 75\% quantiles of \(R_{\text{oos}}^{2}\) across competing methods.}\label{tab:r2_results}
\end{table}

\section{Discussions}

This work proposes a source-data-free transfer learning framework for the SIM. By employing moment estimators derived from a generalized Stein's lemma and a multilayer perceptron (MLP), our method mitigates the dual challenges of model expressivity and source-data accessibility in statistical transfer learning.

\subsection{On the Notion of Transferability}

We seek to clarify and extend the notion of transferability. Specifically, even when entire models are not directly comparable across domains, the similarity between certain components of the source model and those of the target model can still be leveraged to improve estimation on the target domain. Similar to our work,~\citet{Tian02102023} note that their methods also apply to cases where different domains have their own link functions; however, they work within the framework of generalized linear models and impose more restrictive conditions on those link functions. Unlike their GLM-based framework, our work adopts SIM, in which we exploit similarity among indices while abstracting away variations in link functions. The similarity we exploit lies not in the parameter values themselves, but in the one-dimensional embedding space defined by the index. This is the key insight of our work. Consequently, a natural extension of this work is to generalize the proposed framework to the multi-index model, which would encode a higher-dimensional latent space and thereby broaden its applicability to more complex, nonlinear learning problems.

\subsection{Choices of Methods on Source Domains}

On the source domains, we adopt a moment‑based estimator derived from Stein’s lemma. 

Compared with classical sufficient dimension reduction methods such as sliced inverse regression (SIR)~\citep{Li01061991} and minimum average variance estimation (MAVE)~\citep{xia2002}, the Stein‑type estimator imposes weaker conditions on both the covariate distribution and the link function. In particular, it does not require the linearity condition (needed by SIR) nor the second‑order smoothness of the link function (needed by MAVE).

An alternative approach on the target domain is to train a deep neural network (DNN) directly using target data. Once the DNN is obtained, the index can be estimated using the method described in Section~\ref{sec:cm}. However, even when the DNN approximates the conditional expectation accurately, this estimated index can be inaccurate~\citep{tian2025nonlinear}.

For an SIM, the gradient of the conditional expectation is proportional to the index. One might therefore consider using the normalized gradient of the fitted DNN as an alternative index estimator. This gradient‑based estimator is supervised, while Stein’s score is unsupervised; thus, the former could be more efficient. However, the theoretical analysis of DNN‑based gradient estimation is considerably more involved, and establishing its finite sample error rates remains challenging.

The moment‑based Stein estimator also enables straightforward incorporation of $\ell_1$ or even $\ell_0$ sparse regularizations, which is particularly useful for high‑dimensional variable selection on the index coefficients. We leverage the hard-thresholding estimator for the source domains, since the source data are assumed to be sufficiently ample and the estimator is unbiased. 

Feasibility of the proposed method hinges on the accuracy of the estimated score functions, which may require a sufficiently large sample size. When labeled source data are limited, unlabeled data can be incorporated to improve score estimation, as this is a purely unsupervised problem.

\subsection{Choices of Methods on the Target Domain}

The theoretical and practical bottleneck of the proposed target index estimators lies in the difficulty of estimating Stein's scores in ultra-high-dimensional problems of the target domain, where \(d\) is much larger than \(n_0\) (e.g., \(d/n_0 \to \infty\)). Although the current framework possesses the ability to handle high-dimensional problems, it is more suitable for regimes where \(d\) and \(n_0\) are of the same order, i.e., \(d/n_0 \to c\) for some constant \(c > 1\). In such settings, the score estimator can be sufficiently accurate to provide a reasonably informative direction for the target index. Several methods could mitigate the issue of dimensionality. First, if prior knowledge suggests that the link function is ``close'' to linear, the Lasso estimator can serve as an alternative to the score‑based estimator. Second, more advanced neural score estimators, such as NDSM~\citep{birrell2025}, could be employed. However, theoretical analyses of their convergence properties would be much more complicated.

\subsection{About Privacy}

As a natural extension of the proposed framework, one may consider integrating differential privacy (DP) guarantees. In the current work, we do not explicitly inject noise to achieve DP, as our primary focus is on statistical accuracy of transfer learning under data inaccessibility rather than on formal privacy preservation. However, it is worth noting that if the estimated source indices are perturbed with appropriately calibrated noise, differential privacy can be readily attained. This direction would yield a differentially private version of our transfer learning procedure, allowing for rigorous privacy guarantees while retaining the benefits of index‑based transfer. We leave this investigation to future work.

\section*{Disclosure statement}\label{disclosure-statement}

The authors report there are no competing interests to declare.

\section*{Data Availability Statement}\label{data-availability-statement}

All numerical studies were conducted using Python code. The data and Python code are available at \url{https://gitee.com/yetnenu/multi-source-transfer-learning-of-sparse-single-index-models}.
\newpage

\bibliographystyle{apalike}
\bibliography{bibliography.bib}

\end{document}